\documentclass[11pt,a4, oneside]{article}
\usepackage{amsthm,amsmath,latexsym,amssymb,amsfonts,amscd}
\usepackage{graphics,lscape,fancyhdr,array,stmaryrd,euscript}
\usepackage{ifthen,empheq,slashed}
\usepackage{color}
\numberwithin{equation}{section}
\usepackage{setspace}
\usepackage{cite}
\usepackage[hang,flushmargin]{footmisc} 
\usepackage[citecolor=purple,linkcolor=blue,urlcolor=blue,colorlinks=true,filecolor=green,
pdfstartview=FitV, 
pdftitle={},
pdfauthor={Mojtaba Najafizadeh},
pdfsubject={},
pdfkeywords={},
pdfpagemode=None,
bookmarksopen=true,
hyperfootnotes=None]{hyperref}
\usepackage{makecell}
\usepackage{geometry}
\newgeometry{left=0.8in,right=0.8in,top=.8in,bottom=1.1in}
\usepackage[most]{tcolorbox}
\usepackage{enumitem}
\usepackage{footnotebackref}

\def\be {\begin{equation}}
	\def\ee {\end{equation}}
\def\bea {\begin{eqnarray}}
	\def\eea {\end{eqnarray}}
\def\bc {\begin{center}}
	\def\ec {\end{center}}
\def\bg {\begin{align}}
	\def\eg {\end{align}}
\def\bi {\begin{itemize}}
	\def\ei {\end{itemize}}

\def\le {\left}
\def\ri {\right}
\def\p {\partial}

\def\vs {\vspace}

\def\c  {\cdot}
\def\a  {\alpha}
\def\b  {\beta}
\def\g  {\gamma}
\def\d  {\delta}

\def\e  {\eta}

\def\m  {\mu}
\def\n  {\nu}
\def\r  {\rho}
\def\w {\omega}
\def\dw{\wb}
\def\dww{\wb^{\,2}}

\def\wdx {(\omega\cdot\p)}
\def\dwdx {(\wb\cdot\p)}

\def\wb {\overbar\omega}
\def\wbb {\overbar\omega^{\,2}}
\def\ws{\omega\hspace{-8pt}\slash\hspace{3pt}}
\def\wbs{\wb\hspace{-8pt}\slash\hspace{3pt}}
\def\es{\eta\hspace{-5pt}\slash\hspace{0pt}}
\def\eb {\bar\eta}
\def\ebs{\eb\hspace{-5pt}\slash\hspace{0pt}}
\def\ds{\partial\hspace{-6pt}\slash\hspace{1pt}}
\def\ep{\varepsilon}
\newcommand{\overbar}[1]{\mkern 1.5mu\overline{\mkern-1.5mu#1\mkern-1.5mu}\mkern 1.5mu}
\def\1{_{_1}}
\def\2{_{_2}}

\def\suz{\underset{s=0}{\overset{\infty}{\textstyle\sum}}}
\def\suo{\underset{s=1}{\overset{\infty}{\textstyle\sum}}}

\begin{document}

	\hfill
	\begin{flushright}
		{IPM/P-2021/42}
	\end{flushright}
	\vskip 0.08\textheight
	\begin{center}

		{\bfseries \LARGE{Off-shell Supersymmetric Continuous Spin Gauge Theory\vspace{1cm}}} \\
		
		\vskip 0.04\textheight

		Mojtaba \textsc{Najafizadeh} 
		

		
		\vskip 0.01\textheight

		\vspace*{5pt}
		{\em 
			School of Physics, Institute for Research in Fundamental Sciences (IPM), \\
			P.O.Box 19395-5531, Tehran, Iran}
		
		
		\vskip 0.01\textheight
		
		\href{mailto:mnajafizadeh@ipm.ir}{mnajafizadeh@ipm.ir}
		
		\vskip 0.05\textheight

		{\bf Abstract }
		
	\end{center}
	\begin{quotation}
		We construct, for the first time, an off-shell supersymmetric continuous spin gauge theory in 4-dimensional Minkowski spacetime, in both constrained and unconstrained Lagrangian formulations. As an extension to the on-shell description \cite{Najafizadeh:2019mun}, we employ an auxiliary field to close the algebra of supersymmetry transformations off-shell. The 4d $\mathcal{N}=1$ massless continuous spin supermultiplet is then denoted by $(\mathrm{\Phi}, \mathrm{H} \,; \mathrm{\Psi})$, comprised of a dynamical and a non-dynamical complex scalar continuous spin gauge fields $\mathrm{\Phi}$ and $\mathrm{H}$, as well as a Dirac continuous spin gauge field $\mathrm{\Psi}$. In particular, we demonstrate that the off-shell continuous spin supermultiplet, in a limit, reproduces off-shell supersymmetry transformations of the known scalar supermultiplet $(\,{\scriptstyle 0}\,,\, {\scriptstyle 1/2}\,)$, all integer-spin supermultiplets $(\,s\,,\, s\, {\scriptstyle +\, 1/2}\,),\, s\geqslant1$, and all half-integer spin supermultiplets $(\, s\, {\scriptstyle -\, 1/2}\,,\, s \,\,),\, s\geqslant1$.
	\end{quotation}


	\newpage
	\tableofcontents
	\newpage
	\section{Introduction}
	Elementary particles propagating on 4-dimensional Minkowski spacetime have been classified a long time ago by Wigner, using the unitary irreducible representations (UIRs) of the Poincar\'e group $ISO(3,1)$ \cite{Wigner} (See also \cite{Bekaert:2006py} for more details in any dimension). According to Wigner's classification, massive particles are determined by representations of the rotation group $SO(d-1)$, while massless particles are divided into two classifications, the ``helicity'' and ``continuous spin''\,\footnote{Also known as ``infinite spin'' representation in the literature.} representations. Indeed, all known massless particles (helicity particles) possessing a {\it finite} number of physical degrees of freedom per spacetime point are characterized by representations of the Euclidean group $E_{d-2}=ISO(d-2)$. On the other hand, exotic massless particles, called continuous spin particles (CSPs), having an {\it infinite} number of physical degrees of freedom per spacetime point are determined by representations of the short little group $SO(d-3)$, which is the little group of $E_{d-2}$\,\cite{Brink:2002zx}. We note that although continuous spin representations are known for many decades, corresponding continuous spin particles (if any) have not yet been observed in colliders with current energy scales.

	\vspace{.5cm}

	Apart from the method of induced representation, there is another method for classifying UIRs of the Poincar\'e group by making use of the eigenvalues of the Casimir operators; the quadratic Casimir operator $\mathcal{C}_2:=P^2$ (the square of the momentum $P_\m$), and the quartic Casimir operator $\mathcal{C}_4:=W^2$ (the square of the Pauli-Lubanski vector ${W}^\m:= \frac{1}{2} \, \epsilon^{\m\n \rho\sigma} \, {P}_\n \, {J}_{\rho\sigma}$). Accordingly, massless representations, dividing into the helicity and continuous spin representations, are characterized by the following eigenvalues:
	\be
	\hbox{helicity rep.}\,\equiv\,\left\{
	\begin{aligned}
		P^2&=0 
		\\
		W^2&=0
	\end{aligned}
	\right.\,,	
	\qquad\qquad
	\hbox{continuous spin rep.}\,\equiv\,\left\{
	\begin{aligned}
		P^2&=0
		\\
		W^2&=\m^2
	\end{aligned}
	\right.\,.	\nonumber
	\ee
	We note that, in this approach, although the helicity representations have vanishing Casimir operator eigenvalues, $W^2=0$ can be satisfied if one considers $W^\m=-\,h\,P^\m$ where $h$ represents a proportionality factor (operator). This factor can be then determined by $h=-\,{ W^0/P^0}={ \vec{S}\c \hat{P}}$, which is nothing but the helicity operator. Therefore, the helicity representations are labeled by their helicity $h$ and the helicity states do not mix under Lorentz boosts. On the other hand, the continuous spin representations having a non-vanishing eigenvalue are labeled by a dimensionful parameter, called continuous spin parameter $\m$ (a real parameter with the dimension of a mass). Notice that, unlike the helicity states, the continuous spin states are mixed under Lorentz boosts so as the degree of mixing is controlled by $\m$. In the ``helicity limit'', i.e. $\m\rightarrow0$, the continuous spin representation becomes reducible and decomposes into the direct sum of all helicity representations.

	\vspace{.4cm}
	
	About 75 years after Wigner's classification, the first covariant action principles for the bosonic \cite{Schuster:2014hca} and the fermionic \cite{Najafizadeh:2015uxa} continuous spin particles in 4-dimensional flat spacetime were constructed in 2014 and 2015 respectively. The method of obtaining such actions was explained in \cite{Bekaert:2017xin}. In the helicity limit $\m\rightarrow 0$, these actions reproduce respectively the bosonic \cite{Segal:2001qq} and the fermionic \cite{Najafizadeh:2018cpu} higher spin actions in flat spacetime. In particular, these two action principles \cite{Schuster:2014hca,Najafizadeh:2015uxa} have a simple form with no constraints on the gauge fields and gauge parameters. In this sense, one may refer to these as unconstrained Lagrangian formulations of the CSP theory along with the terminology used in the higher spin theory, see references in \cite{Najafizadeh:2018cpu, Najafizadeh:2020moz}.

	\vspace{.4cm}
	
	Afterwards, a constrained Lagrangian formulation \`a la Fronsdal was established for both bosonic \cite{Metsaev:2016lhs} and fermionic \cite{Metsaev:2017ytk} continuous spin fields, in $d$-dimensional (A)dS spacetime, where the gauge fields and gauge parameters are constrained. The detailed procedure of obtaining these actions in $d$-dimensional flat spacetime can be found in \cite{Najafizadeh:2017acd}. In the flat spacetime limit ($\Lambda\rightarrow 0$) and then the helicity limit ($\m\rightarrow 0$), these actions reproduce respectively the Fronsdal \cite{Fronsdal:1978rb} and the Fang-Fronsdal \cite{Fang:1978wz} actions.
	
	\vspace{.5cm}
	
\begin{figure}[h] 
		\centering
		\includegraphics[scale=.20]{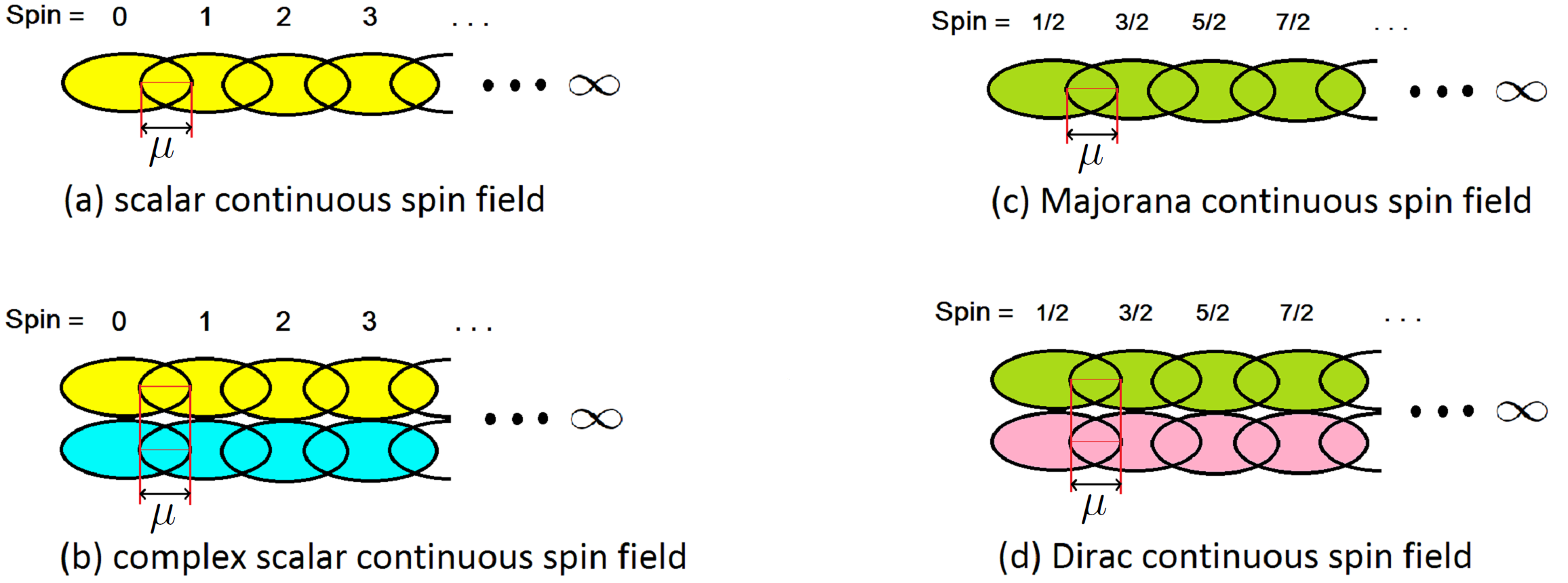}
		\caption{This figure illustrates: (a) the real scalar $\phi$, (b) the complex scalar $\Phi\,(=\phi+i\,\phi')$, (c) the Majorana spinor $\psi$, and (d) the Dirac spinor $\Psi\,(=\psi+i\,\psi')$ continuous spin gauge fields. Remark that each field (a)-(d) describes an ``elementary massless particle'', called continuous spin particle, with an infinite number of physical degrees of freedom per spacetime point. The figure naively shows the continuous spin states (colored ellipses) are mixed under Lorentz boosts which $\m$ controls the degree of mixing. When $\m$ vanishes, the continuous spin gauge fields (a)-(d) decompose into the direct sum of all helicity fields.}
		\label{fig1}
	\end{figure}

	The two above mentioned Lagrangian formulations of the continuous spin theory, i.e. unconstrained and constrained, have recently been supersymmetrized in \cite{Najafizadeh:2019mun} where the algebra of the supersymmetry (SUSY) transformations for the 4d $\mathcal{N}=1$ continuous spin supermultiplet in Minkowski spacetime was closed on-shell. For a better understanding of the supersymmetric continuous spin gauge theory, let us first look at all possible CSP fields in 4 dimensions. From the group/field-theoretical point of view, in 4 dimensions, there exist only two types of continuous spin representations/fields:
	\begin{itemize}[font=\bfseries]
		\item Bosonic continuous spin gauge field whose spectrum includes all integer ($s=0,1,\ldots,\infty$) helicities. This can be itself of two types, a real scalar or a complex scalar.
		\item Fermionic continuous spin gauge field whose spectrum comprises all half-integer ($s={\scriptstyle \frac{1}{2}},{\scriptstyle \frac{3}{2}},\ldots,\infty$) helicities. This also might be of two kinds, a real (Majorana) spinor or a complex (Dirac) spinor.
	\end{itemize}
	Therefore, there are no CSP fields of type the so-called vector, vector-spinor, and so on. In other words, denoting the bosonic and the fermionic CSP fields as the spin-$0$ and the spin-$\frac{1}{2}$ CSP fields \cite{Metsaev:2021zdg}, there would be no spin-$s$ continuous spin field with $s\geqslant1$ in four dimensions. To shed light on the continuous spin supermultiplet, let us visualize each field as shown in Fig.\,\ref{fig1}.
	\vspace{.5cm}	
	
\begin{figure}[h]
		\centering
		\includegraphics[scale=.25]{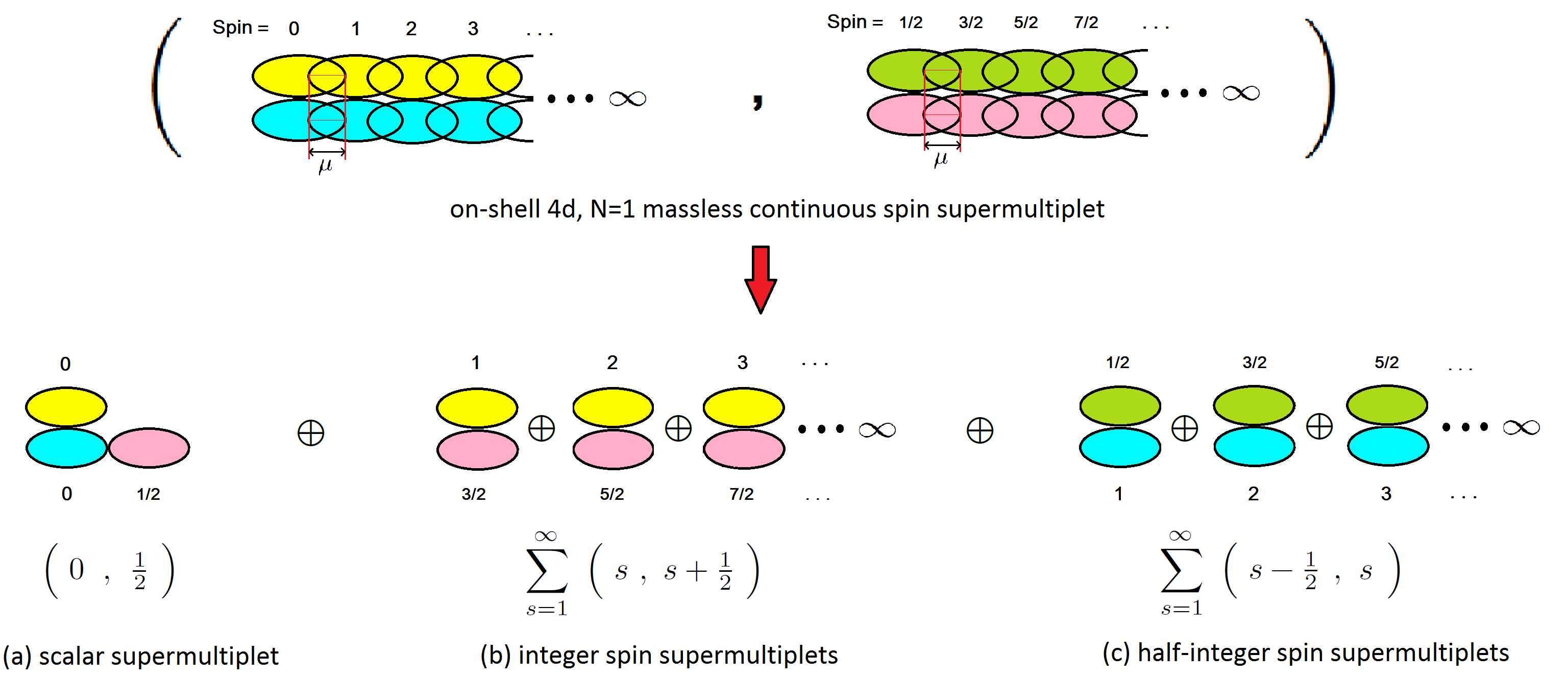}
		\caption{This figure illustrates, when $\m$ vanishes, the continuous spin supermultiplet, comprising of the complex scalar and the Dirac continuous spin gauge fields, decomposes into a direct sum of all known helicity supermultiplets \cite{Najafizadeh:2019mun}.}
		\label{fig2}
	\end{figure}

	As mentioned, a CSP field has infinite number of physical degrees of freedom, hence, the equality of the number of bosonic and fermionic degrees of freedom in a CSP supermultiplet looks like meaningless. Therefore, in 4-dimensional flat spacetime, there would be in principle four possibilities for the $\mathcal{N}=1$ supermultiplet containing of a bosonic (real or complex) and a fermionic (Majorana or Dirac) continuous spin fields. Among these possibilities, we observed \cite{Najafizadeh:2019mun} that the only choice which is consistent with supersymmetry expectations is the case including a complex scalar CSP and a Dirac CSP fields. This selection, in addition, can give us an important result from the helicity limit perspective, under which all known supermultiplets should be recovered. In fact, we demonstrated that, when $\m$ vanishes, on-shell supersymmetry transformations of the continuous spin supermultiplet results in those of all known supermultiplets, such as the scalar supermultiplet $(\,{\scriptstyle 0}\,,\, {\scriptstyle 1/2}\,)$; all integer spin supermultiplets $(\,s\,,\, s\, {\scriptstyle +\, 1/2}\,),\, s\geqslant1$; and all half-integer spin supermultiplets $(\, s\, {\scriptstyle -\, 1/2}\,,\, s \,\,),\, s\geqslant1$. This result presented in \cite{Najafizadeh:2019mun} can be visualized on Fig.\,\ref{fig2}.
	
	\vspace{.5cm}
         
        Apart from the above discussion, other supersymmetric formulations of the CSP theory can be found in \cite{Brink:2002zx, Zinoviev:2017rnj, Buchbinder:2019iwi, Buchbinder:2019kuh, Buchbinder:2019esz}. Furthermore, a number of papers have studied other aspects of the continuous spin gauge theory in different approaches \cite{Khan:2004nj}-\!\!\cite{Buchbinder:2021bgv}. For instance:
	\begin{itemize}
		\item  It was shown the continuous spin representation can be obtained either from the massive representations by taking a suitable limit \cite{Khan:2004nj} (see also \cite{Bekaert:2005in, Najafizadeh:2017acd}), or from contraction of the conformal algebra \cite{Khan:2021dgj}. 
		
		\item Although the helicity and the continuous spin representations are massless, only the helicity representation of the Poincar\'e algebra has a conformal extension, there is no conformal extension to the continuous spin representation \cite{Khan:2004nj} (see also \cite{Bekaert:2009pt}).
		
		\item Since an interacting theory is more favored, possible interactions of continuous spin particle with matter were investigated in \cite{Metsaev:2017cuz,Bekaert:2017xin}, while interactions of continuous spin tachyon\,\footnote{Also known as massive continuous spin particle.} were studied in \cite{Rivelles:2018tpt,Metsaev:2018moa}.

		\item Although CSP fields are massless, the Dirac continuous spin field equation can not be decoupled into two Weyl equations, as what happens in the massless Dirac spin-${\frac{1}{2}}$ field. In fact, continuous spin parameter $\m$ plays a similar roll as mass in the massive Dirac spin-${\frac{1}{2}}$ field equation \cite{Najafizadeh:2019mun}. 
			
	\end{itemize}
	
	\vspace{.5cm}
	
	
	
	
	In this work we construct, for the first time, off-shell formulations of the supersymmetric continuous spin gauge theory for the 4d $\mathcal{N}=1$ supermultiplet at the level of action principles. We note that, at the level of Wigner's wavefunctions, a superfield (off-shell) description was first discussed in \cite{Buchbinder:2019esz}, while other formulations studied only on-shell description \cite{Zinoviev:2017rnj, Buchbinder:2019iwi, Najafizadeh:2019mun, Buchbinder:2019kuh}. To construct an off-shell Lagrangian formulation, we extend on-shell descriptions presented in \cite{Najafizadeh:2019mun}. For this purpose, we introduce an auxiliary field, denoted by $\mathrm{H}$, and find that it should be a complex scalar continuous spin gauge field, which is added into the on-shell supermultiplet, i.e.
	\be 
	\hbox{Off-shell 4d, $\mathcal{N}=1$ continuous spin supermultiplet} \qquad \Rightarrow \qquad \Big(~\Phi~,~\mathrm{H}~;~\Psi~\Big)\label{CSP super}\,.\nonumber
	\ee  
Therefore, by taking into account the unconstrained \cite{Schuster:2014hca, Najafizadeh:2015uxa} and constrained \cite{Metsaev:2016lhs, Metsaev:2017ytk} Lagrangian formulations, we introduce appropriate auxiliary actions corresponding to each formulation, and construct two off-shell Lagrangian formulations for the SUSY CSP theory. We then provide two set of supersymmetry transformations leaving two related off-shell actions invariant. It is shown the algebra of each supersymmetry transformations is closed off-shell, i.e. no equations of motion are applied. Afterwards, by rephrasing off-shell Lagrangian formulation in terms of real continuous spin gauge fields (i.e. in terms of the real scalar and the Majorana CSP fields, Fig.\,\ref{fig1}), and then by taking the helicity limit ($\m\rightarrow0$), we arrive at the so-called ``off-shell supersymmetric higher spin theory''. We can then recover all off-shell Lagrangian formulations of the known supermultiplets. Our results are summarized in table \ref{Table1}.
	\begin{table}[h]
		\centering
		\caption{Off-shell 4d, $\mathcal{N}=1$ massless supermultiplets in flat space, obtained (reproduced) in this paper}
		\vspace{.15cm}
		{\begin{tabular}{|c||c|c|c|c|c|}
				\hline
				{\footnotesize Supermultiplets :} & \makecell{{\footnotesize Continuous spin} \\ {\footnotesize unconstr./constr.} } &  \makecell{ {\footnotesize Continuous spin} \\ {\footnotesize / Higher spin}} & {\footnotesize Scalar\,(chiral)} & \makecell{{\footnotesize Integer} \\ {\footnotesize spins $s\geqslant1$}}  & \makecell{ {\footnotesize Half-integer} \\{\footnotesize  spins $s\geqslant1$}} \\\hline\hline
				\small{ Gauge fields :} & complex & real & real & real & real \\\hline
				\footnotesize{(bosons ; fermions):} & $(\,\Phi,\, \mathrm{H}\, \,;\, \Psi\,)$ & $(\,\phi\,,\,\phi'\,,\,h\,,\,h' \,; \psi\,,\,\psi')$ & $(\phi,\phi',h\,,h' \,; \psi)$ & $(\phi_s,h_s \,; \psi_s)$ & $(\phi'_s,h'_s \,;\psi'_{s-1})$ \\\hline
				\small{\# off-shell d.o.f :} & $(\infty , \infty \,; \infty)$ & $(\infty,\infty,\infty, \infty \,;\infty, \infty)$  & $(1\,,\,1\,,\,1\,,\,1\,;4\,)$ & $(\,2~,~2\,\,;\,4\,\,)$ & $(\,2~,~2\,\,;\,\,\,~4~\,\,)$ \\\hline
				\small{\# on-shell d.o.f :} & $(\infty , \,0\,\, \,; \infty)$ & $(\infty,\infty,\,0\,\,, \,0\,\, \,;\infty, \infty)$ & $(1\,,\,1\,,\,0\,,\,0\,;2\,)$ & $(\,2~,~0\,\,;\,2\,\,)$ & $(\,2~,~0\,\,;\,\,\,~2~\,\,)$ \\\hline
				\small{SUSY actions :} & \,\eqref{off susy u action}\, / \eqref{susy off const}  & \eqref{real csp action} / \eqref{real hs action} & \eqref{wz action} & \eqref{integer sup-} & \eqref{half-integer sup-} \\\hline
				\small{SUSY trans.\,:} & \eqref{susy tr} / \eqref{susy tr cons}  & \eqref{real susy tr} / \eqref{real susy tr hs} & \eqref{wz susy tra} & \eqref{s,s+1/2} & \eqref{s-1/2,s} \\\hline
		\end{tabular}}
		\label{Table1}
	\end{table}

	
	
	
	This paper is organized as follows. In section \ref{unconstrained formu}, we will introduce an auxiliary action to construct the off-shell supersymmetric action for unconstrained formulation. We then provide supersymmetry transformations leaving the action invariant and check the closure of the SUSY algebra. In section \ref{constrained formu}, we will define a related auxiliary action to establish the off-shell supersymmetric action for constrained formulation. We then find supersymmetry transformations and close the algebra off-shell. In section \ref{reformulation}, we rephrase our constrained Lagrangian formulation in terms of real CSP fields. In section \ref{Extraction}, we take the helicity limit ($\m\rightarrow0$) and reproduce all know off-shell supermultiplet results, i.e. the scalar supermultiplet and all higher spin supermultiplets. The conclusions are displayed in section \ref{conclu}. In appendices; we present our conventions in the appendix \ref{conven}. The appendix \ref{invariance} demonstrates invariance of the off-shell action under SUSY transformations. Useful relations concerning supersymmetry and so on will be presented in the appendix \ref{usef}.

	\section{Unconstrained Lagrangian formulation}\label{unconstrained formu}
	On-shell supersymmetric action of the continuous spin gauge theory was given by a sum of the bosonic and the fermionic CSP actions \cite{Najafizadeh:2019mun}. The bosonic part included the Schuster-Toro action \cite{Schuster:2014hca} (up to a partial integration), in which the scalar continuous spin gauge field $\Phi$ was considered to be complex. The fermionic part was given by \cite{Najafizadeh:2015uxa} in which the fermionic continuous spin gauge field $\Psi$ was taken to be a Dirac spinor. Now, in order to construct an off-shell formulation for the theory, we have to add an auxiliary action to the bosonic part of the on-shell system. Such an auxiliary action describes an auxiliary scalar continuous spin gauge field (denoted by $\mathrm{H}$) which is considered to be complex. The auxiliary field $\mathrm{H}$ is non-dynamical and has infinite degrees of freedom per spacetime point. 
	
	
	\subsection{Off-shell supersymmetric continuous spin action}

	In unconstrained Lagrangian formulation, we find the off-shell supersymmetric continuous spin action as a sum of three actions
	\vspace{0cm}
	\begin{subequations} 
		\label{off susy u action}
		\begin{tcolorbox}[ams align, colback=white!98!black]	
			\hspace{0cm}	{S}_{_{\hbox{{\tiny \,SUSY}}}}^{^{\hbox{{\,\tiny CSP}}}}
			&=\int d^4x\, d^4\e\,\,\delta'(\e^2-1)\,\Big[\,\Phi^\dagger(x,\e)~\Bbb{B}~\Phi(x,\e)~+~\mathrm{H}^\dagger(x,\e)~\mathrm{H}(x,\e)~+~\overbar{\Psi}(x,\e)\,(\es-1)\,\Bbb{F}~\Psi(x,\e)\,\Big],
		\end{tcolorbox}
		\hspace{-.7cm}
		 where the bosonic $\Bbb{B}$ and the fermionic $\Bbb{F}$ operators are defined as 
		\begin{align}
			\Bbb{B}&:=\Box - (\e\c\p)(\eb\c\p+\m\,)+\tfrac{1}{2}\,(\e^2-1)(\eb\c\p+\m\,)^2\,, \label{operator B} \\[5pt]
			\Bbb{F}&:=\ds - (\es+1)(\eb\c\p+\m\,)\,. \label{operator F}
		\end{align}
	\end{subequations}
	In above action, $\m$ is continuous spin parameter\,\footnote{\label{foot}\,Without loss of generality, one may flip the sign of $\m$ and define another bosonic and fermionic operators in \eqref{off susy u action}.}, $\e^{\a}$ is a 4-dimensional auxiliary Lorentz vector localized to the unit hyperboloid $\e^2=1$ of one sheet, $\g^\a$ are the 4-dimensional Dirac gamma matrices, and $\d'$
	is the derivative of the Dirac delta function with respect to its argument, i.e. $\delta'(a)=\frac{d}{da}\,\delta(a)$. Other associated quantities in the action \eqref{off susy u action}, such as $\eb_\n$, $\p_\n$, the d'Alembertian operator $\Box$, the Dirac adjoint $\overline{\Psi}$, and the Dirac slash notations $\es$, $\ds$ are introduced in \eqref{def1},\eqref{def2}, and \eqref{def3} respectively. The action \eqref{off susy u action} is Hermitian, $S^\dagger=S$, with respect to the Hermitian conjugation rules \eqref{Hermitian con 1}. In what follows, let us describe each part of the action \eqref{off susy u action} separately in detail.
	
	\subsubsection{Complex scalar continuous spin action}
	
	
	
	
	Bosonic part of the action \eqref{off susy u action}, containing a dynamical field $\Phi$, is given by the complex scalar continuous spin action
	\be 
	S\,[\,\Phi\,]= \int d^4x\, d^4\e~\Phi^\dagger(x,\e)~\delta'(\e^2-1)\,\Bbb{B}~\Phi(x,\e) \,, \label{s-phi}
	\ee   
	where $\Bbb{B}$ was defined in \eqref{operator B}. This action is indeed the Schuster-Toro action \cite{Schuster:2014hca} (up to a total derivative) in which the field has been complexified, and thus, an overall factor of $\tfrac{1}{2}$ has been dropped\,\footnote{Note that the presented action in \cite{Schuster:2014hca} was written in the mostly minus signature for the metric, while the one here \eqref{s-phi} is considered in the mostly plus signature.}. The action \eqref{s-phi} is invariant under two gauge transformations
	\begin{subequations} 
		\begin{align}
			\delta\, \Phi (x,\e)&= \big[    \,\e \cdot \p  -  \tfrac{1}{2}\, (\e^2-1  )   (\eb \cdot \p  +\m\,  ) \big] \,\xi\1 (x,\e)\,,\label{gt b 1} \\[5pt]
			\delta\, \Phi (x,\e)&=(\e^2 - 1 )^2 \, {\xi\2}(x,\e)\,,\label{gt b 2}
		\end{align}
	\end{subequations} 
	where $\xi_1, \xi_2$ are two arbitrary complex gauge transformation parameters. The complex scalar continuous spin field $\Phi$ and two complex gauge transformation parameters $\xi_i \,(i=1, 2)$ are unconstrained and introduce respectively by the generating functions
	\be 
	\Phi(x,\e)=\suz\,\tfrac{1}{s!}~\e^{\m_1} \dots \e^{\m_s}~\Phi_{\m_1 \dots \m_s}(x)\,, \qquad \qquad
	\xi_i(x,\e)=\suo\,\tfrac{1}{(s-1)!}~\e^{\m_1} \dots \e^{\m_{s-1}}~\xi_{i\,\m_1 \dots \m_{s-1}}(x)\,, \label{Phi, xi}
	\ee 
	where $\Phi_{\m_1 \dots \m_s}$ denotes a collection of totally symmetric complex tensor
	fields of all integer rank $s$, and $\xi_{i\,\m_1 \dots \m_{s-1}}$ stands for all totally symmetric complex tensor gauge transformation parameters of integer rank $s-1$. By varying the action \eqref{s-phi} with respect to the gauge fields $\Phi^\dagger$ and $\Phi$, one can obtain two independent equations of motion. For the gauge field $\Phi$, the continuous spin equation of motion reads
	\be 
	\delta'(\e^2-1)\,\Bbb{B}~\Phi(x,\e)=0\,,
	\ee 
	where the bosonic operator $\Bbb{B}$ was introduced in \eqref{operator B}.
	
	\subsubsection{Auxiliary complex scalar continuous spin action}
	
	
	
	
	To construct the off-shell supersymmetric continuous spin action \eqref{off susy u action}, we realized that the following auxiliary action 
	\be 
	S\,[\,\mathrm{H}\,]= \int d^4x\, d^4\e~\mathrm{H}^\dagger(x,\e)~\delta'(\e^2-1)~\mathrm{H}(x,\e) \,, \label{s-H}
	\ee
	have to be added to the bosonic part, in which $\mathrm{H}$ is an auxiliary complex scalar continuous spin gauge field. The action \eqref{s-H} is invariant under the gauge transformation
	\begin{align}
		\delta\, \mathrm{H} (x,\e)=(\e^2 - 1 )^2 \, \upsilon(x,\e)\,,
	\end{align}
	where $\upsilon$ is an auxiliary complex gauge transformation parameter which is an arbitrary function. The auxiliary field $\mathrm{H}$ and gauge transformation parameter $\upsilon$ are both unconstrained and introduce respectively by the generating functions
	\be 
	\mathrm{H}(x,\e)=\suz\,\tfrac{1}{s!}~\e^{\m_1} \dots \e^{\m_s}~\mathrm{H}_{\m_1 \dots \m_s}(x)\,, \qquad \qquad
	\upsilon(x,\e)=\suo\,\tfrac{1}{(s-1)!}~\e^{\m_1} \dots \e^{\m_{s-1}}~\upsilon_{\m_1 \dots \m_{s-1}}(x)\,, \label{H, h}
	\ee
	which have a same structure as \eqref{Phi, xi}, except that they are here auxiliary objects, i.e. $\mathrm{H}_{\m_1 \dots \m_s}$ denotes a collection of totally symmetric auxiliary complex tensor fields of all integer rank $s$, and $\upsilon_{\m_1 \dots \m_{s-1}}$ stands for all totally symmetric auxiliary complex tensor gauge transformation parameters of integer rank $s-1$. By varying the action \eqref{s-H} with respect to the auxiliary gauge fields $\mathrm{H}^\dagger$, $\mathrm{H}$, one may acquire two equations of motion. The equation of motion for the auxiliary continuous spin gauge field $\mathrm{H}$ reads
	\be 
	\delta'(\e^2-1)\,\mathrm{H}(x,\e)=0\,. \label{eom H}
	\ee 
	As one can see, this equation of motion has no kinetic term. Thus, the auxiliary complex scalar continuous spin gauge field $\mathrm{H}$ is a non-dynamical field and consequently does not insert any physical degrees of freedom to the off-shell system. Nevertheless, it is necessary for the full realization of supersymmetry.   
	
	\subsubsection{Dirac continuous spin action}
	
	
	
	The fermionic part of the off-shell SUSY action \eqref{off susy u action} is given by the Dirac continuous spin action \cite{Najafizadeh:2015uxa}
	\be 
	S\,[\,\Psi\,]= \int d^4x\, d^4\e~\overbar{\Psi}(x,\e)\,\delta'(\e^2-1)\,(\es-1)\,\Bbb{F}~\Psi(x,\e) \,, \label{s-psi}
	\ee
	where $\Bbb{F}$ was defined in \eqref{operator F}. The Dirac continuous spin action \eqref{s-psi} is invariant under two spinor gauge transformations
	\begin{subequations}  
		\label{zet}	
		\begin{align}
			\delta \,\Psi(x,\e)&=\big[\,\ds \,(\es +1 \,) - (\eta^2 -1 ) (\eb \cdot \p +\m\,) \,\big]  {\zeta\1}(x,\e)\,, \label{zet 1}\\[5pt]
			\delta \,\Psi(x,\e)&= ( \es -1)\,(\e^2-1)  \, {\zeta\2}(x,\e)=( \es -1) \, {\zeta\2'}(x,\e)  \,,\label{zet 2}
		\end{align}
	\end{subequations} 
	where $\zeta_1$ and $\zeta_2$ (or $\zeta_2':=(\e^2-1)  \, {\zeta\2}$) are two arbitrary spinor gauge transformation parameters. The Dirac continuous spin gauge field $\Psi$ and two spinor gauge transformation parameters $\zeta_i \,(i=1, 2)$ are unconstrained and introduce respectively by the generating functions
	\be 
	\Psi(x,\e)=\suz\,\tfrac{1}{s!}~\e^{\m_1} \dots \e^{\m_s}~\Psi_{\m_1 \dots \m_s}(x)\,, \qquad \qquad
	\zeta_i(x,\e)=\suo\,\tfrac{1}{(s-1)!}~\e^{\m_1} \dots \e^{\m_{s-1}}~\zeta_{i\,\m_1 \dots \m_{s-1}}(x)\,, \label{Psi, zeta}
	\ee 
	where $\Psi_{\m_1 \dots \m_s}$ denotes a collection of totally symmetric Dirac spinor-tensor fields of all half-integer rank $s+\tfrac{1}{2}$, and $\zeta_{i\,\m_1 \dots \m_{s-1}}$ stands for a set of totally symmetric Dirac spinor-tensor gauge transformation parameters of all half-integer rank $s-\tfrac{1}{2}$ (spinor indices are left implicitly). By varying the action \eqref{s-psi} with respect to the fermionic gauge field $\overbar{\Psi}$, one yields the equation of motion for the Dirac continuous spin gauge field $\Psi$ 
	\be
	\delta'(\e^2-1)\,(\es-1)\,\Bbb{F}~\Psi(x,\e)=0\,, \label{eom F}
	\ee
	where the fermionic operator $\Bbb{F}$ was introduced in \eqref{operator F}.
	
	\vspace{.5cm}
	
	We note that in the infinite tower of bosonic \eqref{Phi, xi},\eqref{H, h} and fermionic \eqref{Psi, zeta} spins, every spin state interns only once, and spin states are mixed under the Lorentz boost, so as the degree of mixing is controlled by the continuous spin parameter (see Fig.\,\ref{fig1}). We also notice that in on-shell formulation of the SUSY CSP gauge theory \cite{Najafizadeh:2019mun}, the supersymmetry algebra was closed, by applying the equation of motion, up to a spinor gauge transformation corresponding to \eqref{zet 1}. However, we will see below, in off-shell formulation the algebra would be closed, without using the equation of motion, up to two spinor gauge transformations corresponding to \eqref{zet}. Therefore, the second spinor gauge transformation \eqref{zet 2} will play a crucial role on off-shell closure of the supersymmetry algebra.

	\subsection{Off-shell supersymmetry transformations}
	Now we are in a position to present off-shell SUSY transformations for unconstrained Lagrangian formulation of the CSP gauge theory in Minkowski spacetime for the 4d $\mathcal{N}=1$ supermultiplet. We realize that such an irreducible off-shell supermultiplet should be consist of the complex scalar $\Phi$, auxiliary complex scalar $\mathrm{H}$, and Dirac $\Psi$ continuous spin gauge fields. Therefore, we find that the off-shell supersymmetric continuous spin action \eqref{off susy u action}, including a sum of three actions \eqref{s-phi},\eqref{s-H},\eqref{s-psi}, 
	\be 
	S_{_{\hbox{{\tiny \,SUSY}}}}^{^{\hbox{{\,\tiny CSP}}}}= S\,[\,\Phi\,] \,+\, S\,[\,\mathrm{H}\,]\,+\, S\,[\,\Psi\,]\,,
	\ee 
	is invariant (see appendix \ref{invariance}) under the following supersymmetry transformations 
	\vs{0cm}
	\begin{subequations} 
		\label{susy tr} 
		\begin{tcolorbox}[ams align, colback=white!98!black]
			&\delta\,\Phi(x,\e)\, =\,{\scriptstyle\sqrt{2}}\,\, \bar\ep\,\,L\,\big(\es+1\,\big)\,\Psi(x,\e)\,, \label{susy tr1}\\[8pt]
			&\delta\,\mathrm{H}(x,\e)\, = \,{\scriptstyle\sqrt{2}}\,\,\bar\ep\,\,R\,\big(\es-1\,\big)\,\,\Bbb{F}\,\,\Psi(x,\e) \,,  \label{susy tr2}\\[8pt]
			&\delta\,\Psi(x,\e)\, =\,{\scriptstyle\sqrt{2}}\,\,\Bbb{X}\,\,R\,\,\ep~\Phi(x,\e)\,-\,
			{\scriptstyle\sqrt{2}}\,\,L\,\,\ep\,\mathrm{H}(x,\e)\,. \label{susy tr3}
		\end{tcolorbox}
	\end{subequations}  
	\hspace{-.7cm}
	where the operator $\Bbb{F}$ was introduced in \eqref{operator F}, while the operator $\Bbb{X}$ and chiral projectors $R$, $L$ are defined as
	\be
	\Bbb{X}:=\ds-\tfrac{1}{2}\,\big(\es-1\,\big)\big(\eb\c\p+\m\,\big)\,, \label{bbb X}
	\ee 
	\vspace{.05cm}
	\be 
	L:=\tfrac{1}{2}\,(1+\g^5) \,, \qquad\qquad R:=\tfrac{1}{2}\,(1-\g^5)\,. \label{projections}
	\ee
	In the supersymmetry transformations \eqref{susy tr}, $\ep$ is an arbitrary constant Dirac spinor object that parameterizes the supersymmetry transformations (see \eqref{maj flip} for its property).
	
	\vspace{.5cm}
	
	Let us now calculate commutator of the supersymmetry transformations \eqref{susy tr} acting on the bosonic and fermionic continuous spin gauge fields. We straightforwardly find that the commutator on the continuous spin gauge fields yield
	\begin{subequations}  
		\label{closur}	
		\begin{align}
			[\,\d_1\,,\,\d_2\,]\,\Phi(x,\e)&=\,2\,(\bar\ep\2\,\ds\,\ep\1)\,\Phi(x,\e)\,,\label{b clou1}\\[5pt]
			[\,\d_1\,,\,\d_2\,]\,\mathrm{H}(x,\e)&=\,2\,(\bar\ep\2\,\ds\,\ep\1)\,\mathrm{H}(x,\e)\,,\label{b clou2}\\[5pt]
			[\,\d_1\,,\,\d_2\,]\,\Psi(x,\e)&=\,2\,(\bar\ep\2\,\ds\,\ep\1)\,\Psi(x,\e)\,+\,\hbox{{\small G.T.(\,I\,)}}\,+\,\hbox{{\small G.T.(II)}}\,,\label{f clou}
		\end{align}
	\end{subequations}  
	where $\hbox{{\small G.T.(\,I\,)}}$ and $\hbox{{\small G.T.(II)}}$ stand for two spinor gauge transformations corresponding respectively to \eqref{zet 1} and \eqref{zet 2}, in which spinor gauge transformation parameters are field-dependent, i.e.
	\begin{align}
		\hbox{{\small G.T.(\,I\,)}} & := \big[\,\ds\,(\es+1)-(\e^2-1)\,(\eb\c\p+\m)\,\big]\,\zeta\1(\Psi)\,,\qquad 
		\zeta\1(\Psi)=-\,\g^\n R\,(\bar\ep\2\,\g_\n\,\ep_1)\,\Psi\,,\\[5pt]
		\hbox{{\small G.T.(II)}} &:= (\es-1)~\zeta\2'(\Psi)\,, \qquad \qquad\qquad\qquad\qquad\qquad\,~~\,
		\zeta\2'(\Psi)=\,\tfrac{1}{2}\,\,\g^\n L\,(\bar\ep\2\,\g_\n\,\ep_1)\,\Bbb{F}\,\Psi\,.
	\end{align}
	As a result, we realize that using supersymmetry transformations \eqref{susy tr} the supersymmetry algebra \eqref{closur} is closed off-shell up to two gauge transformations, that is, no equations of motion are used in the calculations, as is expected in an off-shell description. 
	
	\vs{.2cm}
	Remark that if one multiplies supersymmetry transformations \eqref{susy tr} by $\d'(\e^2-1)$ to the left, then the equations of motion \eqref{eom H},\eqref{eom F} will appeare. By applying the equations of motion, supersymmetry transformations \eqref{susy tr} reduce to
	\be 
	\delta\,\Phi\, =\,{\scriptstyle\sqrt{2}}\,\, \bar\ep\,\,L\,\big(\es+1\,\big)\,\Psi\,,
	\qquad \qquad\qquad
	\delta\,\Psi\, =\,{\scriptstyle\sqrt{2}}\,\,\Bbb{X}\,\,R\,\,\ep~\Phi\,,
	\ee  
	where $\Bbb{X}$, $L$, $R$ defined in \eqref{bbb X},\eqref{projections}. These are indeed on-shell SUSY transformations presented in \cite{Najafizadeh:2019mun}.

	\section{Constrained Lagrangian formulation} \label{constrained formu}
	The 4d, $\mathcal{N}=1$ on-shell description of the SUSY CSP gauge theory was given in \cite{Najafizadeh:2019mun}, in which the bosonic and fermionic parts were given by Metsaev actions \cite{Metsaev:2016lhs, Metsaev:2017ytk} in flat spacetime. Similar to the unconstrained formulation \ref{unconstrained formu}, the dynamical fields were considered to be complex scalar $\Phi$ and Dirac $\Psi$ CSP gauge fields, by the difference that gauge fields and parameters were constrained. Here, in order to construct an off-shell description, we add a non-dynamical field $\mathrm{H}$ to the bosonic part. In what follows, we first present the off-shell SUSY action, describe each part of the action, and finally provide SUSY transformations.

	\subsection{Off-shell supersymmetric continuous spin action}
	In constrained Lagrangian formulation, we find the off-shell supersymmetric continuous spin action as a sum of three actions
	\vspace{0cm}
	\begin{subequations}
		\label{susy off const} 
		\begin{tcolorbox}[ams align, colback=white!98!black]	
			\hspace{-.4cm}S_{_{\hbox{{\tiny \,SUSY}}}}^{^{\hbox{{\,\tiny CSP}}}}&=\!\int d^4 x\, \Big[\Phi^\dagger(x,\wb)\,\big(\,\mathrm{B}+\mathrm{B}_\m\,\big)\,{\Phi}(x,\w)\,+\,\mathrm{H}^\dagger(x,\wb)\,\big(\,{\mathrm{B}_{_0}}\,\big)\,\mathrm{H}(x,\w)\,+\,\overline{\Psi}(x,\wb)\,\big(\,\mathrm{F}+\mathrm{F}_\m\,\big)\,\Psi(x,\w)
			\Big]\bigg|_{\w=0},\!\!\!\!\!
		\end{tcolorbox}
		\hspace{-.7cm}
		where the bosonic $\mathrm{B}$, $\mathrm{B}_\m$, $\mathrm{B}_{_0}$ and the fermionic $\mathrm{F}$, $\mathrm{F}_\m$ operators are given by
		\begin{align}
			\mathrm{B}_{~}&:=\Box-\wdx\dwdx+\tfrac{1}{2}\,\wdx^2\dww +\tfrac{1}{2}\, \w^2\dwdx^2-\tfrac{1}{2}\,\w^2\,\Box\,\dww - \tfrac{1}{4}\,\w^2\wdx\dwdx\,\dww \,,\label{B}\\[5pt]
			\mathrm{B}_\m&:=\,\m\,\Big[ \le(\w\c\p-\w^2\dwdx+\tfrac{1}{4}\,\w^2\wdx\,\dww\ri)\tfrac{-\,1}{\sqrt{\smash[b]{2(N+1)}}}+\tfrac{1}{\sqrt{\smash[b]{2(N+1)}}} \le(\wb\c\p-(\w\c\p)\,\wbb+\tfrac{1}{4}\,\w^2\,(\wb\c\p)\,\wbb\ri)\Big]\,, \nonumber\\[5pt]
			&~+\m^2\Big[\,\tfrac{1}{2(N+1)}+\w^2\,\tfrac{1}{8(N+3)}\,\dww+\,\tfrac{1}{4}\,\w^2\,\tfrac{1}{\sqrt{\smash[b]{(N+1)(N+2)}}}+\,\tfrac{1}{4}\,\tfrac{1}{\sqrt{\smash[b]{(N+1)(N+2)}}}\,\dww\,\Big]\,, \label{Bmu} \\[5pt]
			\mathrm{B}_{_0}&:= \, 1-\tfrac{1}{4}\,\w^2\,\wbb\,, \label{B-0}\\[5pt]
			\mathrm{F}_{~}&:=\,-\,\Big[\, \ds-\ws\dwdx-\wdx\,\wbs + \ws\ds\,\wbs + \tfrac{1}{2}\,\ws\wdx\,\dww +\tfrac{1}{2}\,\w^2\dwdx\,\wbs -\tfrac{1}{4}\,\w^2\ds \,\dww \Big]\,, \label{F}\\[5pt]
			\mathrm{F}_\m&:=\,\m\,\,\Big[\tfrac{1}{N+1}\,\le(1-\ws\wbs-\tfrac{1}{4}\,\w^2\dww\ri)+\le(\ws-\tfrac{1}{2}\,\w^2\wbs\ri)\tfrac{-\,1}{\sqrt{\smash[b]{2(N+1)}}}+\tfrac{1}{\sqrt{\smash[b]{2(N+1)}}}\le(\wbs-\tfrac{1}{2}\,\ws\dww\ri)\Big]  \label{Fmu}\,.
		\end{align}
	\end{subequations} 
	As before, $\m$ is continuous spin parameter\,\footnote{Without loss of generality, one may flip the sign of $\m$ and define another bosonic and fermionic operators in \eqref{susy off const}.}, and $\w^\n$ is a 4-dimensional auxiliary Lorentz vector. Other associated quantities in the action \eqref{susy off const}, such as $\wb_\n$, $N$, $\p_\n$, the d'Alembertian operator $\Box$, the Dirac adjoint $\overline{\Psi}$, and the Dirac slash notations $\ws$, $\wbs$, $\ds$ are introduced in \eqref{def1},\eqref{def2}, and \eqref{def3} respectively. The action \eqref{susy off const} is Hermitian (i.e. $S^\dagger=S$) with respect to the Hermitian conjugation rules \eqref{hermitian conjugates 2}. In what follows, we shall describe each part of the action \eqref{susy off const} separately in detail.
	
	\subsubsection{Complex scalar continuous spin action}
	
	Bosonic part of the action \eqref{susy off const}, including a dynamical field $\Phi$, is given by the complex scalar continuous spin action
	\be 
	S\,[\,\Phi\,]= \int d^4x~\Phi^\dagger(x,\wb)\,\big(\,\mathrm{B}\,+\,\mathrm{B}_\m\,\big)\,{\Phi}(x,\w)~\Big|_{\w=0}\,, \label{s-phi-}
	\ee
	where $\mathrm{B}$, $\mathrm{B}_\m$ were introduced in \eqref{B},\eqref{Bmu}. This action is indeed the Metsaev action \cite{Metsaev:2016lhs} in 4-dimensional flat space, in which the gauge field has been considered to be complex, as a result, an overall factor of $\tfrac{1}{2}$ has been removed. The operators $\mathrm{B}$, $\mathrm{B}_\m$, and consequently the action \eqref{s-phi-}, are Hermitian\,\footnote{i.e. $\mathrm{B}^\dagger=\mathrm{B}$, $\mathrm{B}_\m^\dagger=\mathrm{B}_\m$, $(\,S\,[\Phi]\,)^\dagger=S\,[\Phi]$.} with respect to the Hermitian conjugation rules \eqref{hermitian conjugates 2}. 
	
	\vspace{.5cm}
	
	The action \eqref{s-phi-} is invariant under the gauge transformation
	\be 
	\delta\, {{\Phi }} (x,\w)\,=\,\Big(\w \cdot \p \,- \, \m \, \tfrac{1}{\sqrt{\smash[b]{2(N+1)}}} \,+ \, \m ~ \w^2\,\tfrac{1}{\,{2(N+1)}\,\sqrt{\smash[b]{2(N+2)}}\,}\, \Big)\, \chi(x,\w) \,,  \label{Gauge T m}
	\ee 
	where $\chi$ is a complex gauge transformation parameter. Unlike the previous section \ref{unconstrained formu}, here, we deal with constrained formulation. This means that the complex scalar continuous spin gauge field $\Phi$ and complex gauge transformation parameter $\chi$, which are introduced by the generating functions\,\footnote{In \eqref{Phi, xi,w}, $\Phi_{\m_1 \dots \m_s}$ denotes a collection of totally symmetric complex tensor fields of all integer rank $s$, and $\chi_{\m_1 \dots \m_{s-1}}$ stands for all totally symmetric complex tensor gauge transformation parameters of integer rank $s-1$.}  
	\be 
	\Phi(x,\w)=\suz\,\tfrac{1}{s!}~\w^{\m_1} \dots \w^{\m_s}~\Phi_{\m_1 \dots \m_s}(x)\,, \qquad \quad
	\chi(x,\w)=\suo\,\tfrac{1}{(s-1)!}~\w^{\m_1} \dots \w^{\m_{s-1}}~\chi_{\m_1 \dots \m_{s-1}}(x)\,, \label{Phi, xi,w}
	\ee
	obey respectively the double-trace constraint and the trace condition
	\be 
	(\,\wbb\,)^2\,{\Phi}(x,\w)=0 \,, \qquad\qquad\qquad (\,\wbb\,)\,\chi(x,\w)=0\,.\label{trace conditions}
	\ee  
	
	By varying the action \eqref{s-phi-}, with respect to the field $\Phi^\dagger$, one finds the scalar CSP equation of motion 
	\begin{align}
		\big(\,\mathrm{B}+\mathrm{B}_\m\,\big)\,{\Phi}(x,\w)=0\,,\label{eom b}
	\end{align} 
	where $\mathrm{B}$, $\mathrm{B}_\m$ were defined in \eqref{B},\eqref{Bmu}. This equation can be reduced to a simpler form. Indeed, by dropping a factor of $(\,1-\frac{1}{4}\,\w^2\,\dww\,)$ from the left-hand-side of \eqref{eom b}, the scalar continuous spin equation of motion may be written in the following form  
	\begin{align}
		\big(\,\mathcal{B}+\mathcal{B}_\m\,\big)\,{\Phi}(x,\w)=0\,,\label{eom b2}
	\end{align} 
	where (the operator $\mathcal{B}$ is indeed the Fronsdal operator \cite{Fronsdal:1978rb})
	\begin{align}
		\mathcal{B}_{~}&:= \Box \,-\, (\w\c\p)(\wb\c\p)\,+\,\tfrac{1}{2}\,(\w\c\p)^2\,\wbb   \,,\\[5pt]
		\mathcal{B}_\m&:= \mu\, \Big[ \, (\omega \cdot \p )\,
		\tfrac{-\,1}{\sqrt{\smash[b]{2(N+1)}}} \,+\,\tfrac{1}{\sqrt{\smash[b]{2(N+1)}}}\, (\wb \cdot \p)\,-\, \omega^2 \, \tfrac{1}{2(N+1)\sqrt{\smash[b]{2(N+2)}}} \, (\wb \cdot \p)  \nonumber\\
		& ~~~~ ~~~~ - \,(\omega\cdot\p)\,\tfrac{1}{\sqrt{\smash[b]{2(N+2)}}}\,\wbb \,+\, \w^2\,(\w\cdot\p)\, \tfrac{1}{2(N+2)\sqrt{\smash[b]{2(N+1)}}}\,\wbb \,\Big]\nonumber\\
		& +   \m^2 \,\Big[\,\tfrac{1}{2(N+1)} - \,\w^2 \, \tfrac{1}{4(N+1)(N+3)}\,\wbb \,-\,  \,\w^2 \, \tfrac{1}{\sqrt{\smash[b]{2(N+2)}}\,[2(N+1)]^{3/2}} \,+\,  \tfrac{1}{4\sqrt{\smash[b]{(N+2)(N+1)}}}\,  \wbb \,\Big] \,.
	\end{align}	
	We note that, in comparison to the spin-two case, one can refer to \eqref{eom b} as the Einstein-like continuous spin equation of motion, while one may refer to \eqref{eom b2} as the Ricci-like continuous spin equation of motion. In other words, using the Hermitian conjugation rules \eqref{hermitian conjugates 2}, one can check that the bosonic operator in \eqref{eom b} is Hermitian $(\mathrm{B}+\mathrm{B}_\m)^\dagger= \mathrm{B}+\mathrm{B}_\m$, while the one in \eqref{eom b2} is non-Hermitian $(\mathcal{B}+\mathcal{B}_\m)^\dagger\neq\mathcal{B}+\mathcal{B}_\m$.

	\vspace{.5cm}
	Recall that when $\m=0$, an infinite sum of higher spin results should be reproduced. As a result, at $\m=0$ (where $\mathrm{B}_\m,\mathcal{B}_\m=0$), bosonic continuous spin formalism, given by the action \eqref{s-phi-} and its accompanying features \eqref{Gauge T m}-\eqref{eom b2}, will reduce to a sum of the Fronsdal formalism \cite{Fronsdal:1978rb}.

	\subsubsection{Auxiliary complex scalar continuous spin action}
	To build the off-shell supersymmetric continuous spin action \eqref{susy off const}, we found the following auxiliary action 
	\be 
	S\,[\,\mathrm{H}\,]= \int d^4x ~\mathrm{H}^\dagger(x,\wb)\,\big(\,{\mathrm{B}_{_0}}\,\big)\,\mathrm{H}(x,\w)~\Big|_{\w=0} \,, \label{s-H-}
	\ee
	should be added to the bosonic part, in which $\mathrm{H}$ is an auxiliary complex scalar continuous spin gauge field, and $\mathrm{B}_{_0}$ was introduced in \eqref{B-0}. The action \eqref{s-H-} is invariant under the gauge transformation
	\begin{align}
		\delta\, \mathrm{H} (x,\w)= 0\,. \label{GT H}
	\end{align}
	The auxiliary complex scalar continuous spin gauge field $\mathrm{H}$ is introduced by the generating function
	\be 
	\mathrm{H}(x,\w)=\suz\,\tfrac{1}{s!}~\w^{\m_1} \dots \w^{\m_s}~\mathrm{H}_{\m_1 \dots \m_s}(x)\,, \label{generating func H}
	\ee 
	where $\mathrm{H}_{\m_1 \dots \m_s}$ denotes a collection of totally symmetric auxiliary complex tensor fields of all integer rank $s$ mixing under Lorentz boosts, so as the degree of mixing is controlled by the continuous spin parameter $\m$. We note that since we are here working in constrained formulation, the auxiliary gauge field is constrained, i.e. obeys the double-trace condition
	\be 
	(\,\wbb\,)^2\,\mathrm{H}(x,\w)=0 \,. \label{trace cond H}
	\ee 
	If one varies the action \eqref{s-H-}, with respect to the gauge field $\mathrm{H}^\dagger$, one finds the equation of motion 
	\be 
	\mathrm{H}(x,\w)=0\,,\label{eom H-}
	\ee 
	demonstrating that there is no kinetic term in the equation, and thus the auxiliary gauge field $\mathrm{H}$ is non-dynamical.
	
	\vspace{.5cm}
	
	Let us recall that here no continuous spin parameter $\m$ was entered in the auxiliary continuous spin action \eqref{s-H-} (and thus in the equation of motion \eqref{eom H-}) to demonstrate that we have actually a continuous spin gauge theory. Nevertheless, as stated, totally symmetric auxiliary complex tensor fields $\mathrm{H}_{\m_1 \dots \m_s}$, packed into the generating function \eqref{generating func H}, were considered to be mixed under the Lorentz boost, demonstrating that we are dealing with a continuous spin gauge theory. In the helicity limit $\m=0$, which higher spin results are expected to be recovered, the form of the auxiliary action \eqref{s-H-} will remain unchanged. However, the generating function \eqref{generating func H} will be interpreted as a collection of totally symmetric auxiliary complex tensor fields $\mathrm{H}_{\m_1 \dots \m_s}$ of all integer rank $s$, which do not mix under the Lorentz boost. We will use this interpretation later to reproduce higher spin supermultiplets from the continuous spin one.

	\subsubsection{Dirac continuous spin action}

	The fermionic part of the off-shell SUSY action \eqref{susy off const} is given by the Dirac continuous spin action 
	\be 
	S\,[\,\Psi\,]= \int d^4x~ \overline{\Psi}(x,\wb)\,\big(\,\mathrm{F}\,+\,\mathrm{F}_\m\,\big)\,\Psi(x,\w)~\Big|_{\w=0}\,, \label{s-psi-}
	\ee
	in which $\Psi$ is the so-called ``Dirac continuous spin gauge field'', and $\mathrm{F}$, $\mathrm{F}_\m$ were defined in \eqref{F},\eqref{Fmu}. This action is equivalent to the Metsaev action \cite{Metsaev:2017ytk} in 4-dimensional flat spacetime. The action \eqref{s-psi-} is invariant under the gauge transformation
	\be 
	\delta\,{\Psi}(x,\w)=\Big(\,\w\cdot \p \,+\,\m\,\tfrac{1}{\sqrt{\smash[b]{2(N+1)}}} \,-\,\m\,\ws\,\tfrac{1}{2(N+1)(N+2)} \,-\,\m\,\w^2\,\tfrac{1}{\raisebox{.09in}~[\,2(N+2)\,]^{3/2}} \,\Big){\tau}(x,\w)\,,
	\label{Gauge T F 2}
	\ee 
	where $\tau$ is the spinor gauge transformation parameter. The Dirac continuous spin gauge field $\Psi$ and spinor gauge transformation parameter $\tau$ are respectively introduced by the generating functions
	\be 
	\Psi(x,\w)=\suz\,\tfrac{1}{s!}~\w^{\m_1} \dots \w^{\m_s}~\Psi_{\m_1 \dots \m_s}(x)\,, \qquad \quad
	\tau(x,\w)=\suo\,\tfrac{1}{(s-1)!}~\w^{\m_1} \dots \w^{\m_{s-1}}~\tau_{\m_1 \dots \m_{s-1}}(x)\,, \label{Psi, tau,w}
	\ee
	where $\Psi_{\m_1 \dots \m_s}$ denotes a collection of totally symmetric Dirac spinor-tensor fields of all half-integer rank $s+\tfrac{1}{2}$, and $\tau_{\m_1 \dots \m_{s-1}}$ stands for a set of totally symmetric Dirac spinor-tensor gauge transformation parameters of all half-integer rank $s-\tfrac{1}{2}$ (spinor indices are left implicitly). The formulation is constrained, that is, the spinor gauge field $\Psi$ and the spinor gauge transformation parameter $\tau$ obey respectively the triple gamma-trace and the gamma-trace conditions
	\be 
	(\,\wbs\,)^3\,{\Psi}(x,\w)=0 \,, \qquad\qquad\qquad (\,\wbs\,)\,\tau(x,\w)=0\,.\label{trace conditions f}
	\ee  
	By varying the action \eqref{s-psi-} with respect to the fermionic gauge field $\overbar{\Psi}$, one can obtain the equation of motion for the Dirac continuous spin gauge field $\Psi$ 
	\be
	\big(\,\mathrm{F}\,+\,\mathrm{F}_\m\,\big)\,\Psi(x,\w)=0\,, \label{F+Fmu}
	\ee
	where $\mathrm{F}$, $\mathrm{F}_\m$ were introduced in $\eqref{F}$,$\eqref{Fmu}$ respectively. If one removes a factor of $(\,1-\frac{1}{2}\,\ws\,\wbs-\frac{1}{4}\,\w^2\,\dww\,)$ from the left-hand-side of \eqref{F+Fmu}, one can then rewrite the Dirac CSP equation in the following form 
	\begin{subequations}  
		\label{F+Fmu--}	
		\be
		\big(\,\mathcal{F}\,+\,\mathcal{F}_\m\,\big)\,\Psi(x,\w)=0\,, \label{F+Fmu-}
		\ee
		where (the operator $\mathcal{F}$ is the Fang-Fronsdal operator \cite{Fang:1978wz})
		\begin{align}
			\mathcal{F}_{~}&:=-\,\ds+(\w\c\p)\,\wbs\,,\label{F-F operator}\\[5pt]
			\mathcal{F}_\m&:=\m\,\Big[\,\tfrac{1}{N+1}+\ws\,\tfrac{2}{\raisebox{.09in}~[2(N+1)]^{3/2}}+\tfrac{1}{\raisebox{.09in}~[2(N+1)]^{1/2}}\,\wbs
			+\ws\,\tfrac{1}{2(N+1)(N+2)}\,\wbs-\w^2\,\tfrac{1}{\raisebox{.09in}~[2(N+2)]^{3/2}}\,\wbs \,\Big]\,.
		\end{align}
	\end{subequations}  
	Similar to the bosonic case, using \eqref{hermitian conjugates 2}, one may check that the fermionic operator in \eqref{F+Fmu} satisfies $(\mathrm{F}+\mathrm{F}_\m)^\dagger=-\,\g^0\, (\mathrm{F}+\mathrm{F}_\m) \,\g^0$, while the one in \eqref{F+Fmu--} reads $(\mathcal{F}+\mathcal{F}_\m)^\dagger\neq-\,\g^0\, (\mathcal{F}+\mathcal{F}_\m) \,\g^0$.
	
	
	
	\vspace{.5cm}
	
	We note again that, at $\m=0$ (where $\mathrm{F}_\m,\mathcal{F}_\m=0$), fermionic continuous spin formalism, given by the action \eqref{s-psi-} and its associated relations \eqref{Gauge T F 2}-\eqref{F+Fmu--}, will reproduce a sum of the Fang-Fronsdal formalism \cite{Fang:1978wz}.

	\subsection{Off-shell supersymmetry transformations}
	In this section, we discussed constrained formulation and presented the off-shell SUSY CSP action \eqref{susy off const}. At this stage we are ready to provide SUSY transformations in flat spacetime for the off-shell 4d $\mathcal{N}=1$ continuous spin supermultiplet. This supermultiplet should be consist of the complex scalar $\Phi$, auxiliary complex scalar $\mathrm{H}$, and Dirac $\Psi$ continuous spin gauge fields. Therefore, we discover that the off-shell supersymmetric continuous spin action \eqref{susy off const}, including a sum of three actions \eqref{s-phi-},\eqref{s-H-},\eqref{s-psi-}, 
	\be 
	S_{_{\hbox{{\tiny \,SUSY}}}}^{^{\hbox{{\,\tiny CSP}}}}= S\,[\,\Phi\,] \,+\, S\,[\,\mathrm{H}\,]\,+\, S\,[\,\Psi\,]\,,
	\ee 
	is invariant under the following supersymmetry transformations
	\vspace{0cm}
	\begin{subequations}  
		\label{susy tr cons}	
		\begin{tcolorbox}[ams align, colback=white!98!black]
			\d \,\Phi(x,\w) &={\scriptstyle\sqrt{2}}\,\,\bar\ep~L~\mathrm{T}~{\Psi}(x,\w)\,, \label{susy tr cons1}\\[5pt]
			\d \,\mathrm{H}(x,\w)&= {\scriptstyle\sqrt{2}}\,\,\bar\ep~R\,\big[\,\mathrm{Y} \,+\, \mathrm{Y}_\m\,\big]\,\Psi(x,\w)
			\,,\label{susy tr cons2}  \\[5pt]
			\d \,\Psi(x,\w) &={\scriptstyle\sqrt{2}} \,\,\big[\, \mathrm{X} \,+\, \mathrm{X}_\m \,\big]\, R~\ep~\Phi(x,\w)\,-\,
			{\scriptstyle\sqrt{2}} ~\mathrm{Z}~L~\ep~\mathrm{H}(x,\w)\,,\label{susy tr cons3}
		\end{tcolorbox}
	\end{subequations}  
	\noindent where $\ep$ is an arbitrary constant Dirac spinor object, $L$, $R$ are chiral projectors defined in \eqref{projections}, and the operators $\mathrm{T}$, $\mathrm{X}$, $\mathrm{X}_\m$, $\mathrm{Y}$, $\mathrm{Y}_\m$, $\mathrm{Z}$ are introduced by
	\begin{align}
		\hspace{-.2cm}	\mathrm{T}~&:=1\,+\,\ws\,\tfrac{1}{\sqrt{\smash[b]{2(N+1)}}} \,,
		\label{T}\\[5pt]
		\hspace{-.1cm}	\mathrm{X}~&:=\ds - \ws\,\tfrac{1}{2(N+1)}\,(\dw\c\p)+\ws\,\ds\,\tfrac{1}{2(N+1)}\,\wbs - \ws\,(\w\c\p)\,\tfrac{1}{4(N+2)}\,\dww+\tfrac{1}{\sqrt{\smash[b]{2(N+1)}}}\,\le[\,\dwdx\,-\,\ds\,\wbs+ \tfrac{1}{2}\,\wdx\,\dww\,\ri]\,,\label{X}\\[5pt]
		\hspace{-.1cm}	 \mathrm{X}_\m&:=\m\,\Big[~\tfrac{1}{2(N+1)}\,-\,\w^2\,\tfrac{1}{\raisebox{.09in}\,8(N+2)(N+3)}\,\dww\,+\,\tfrac{1}{4}\,\tfrac{1}{\sqrt{\smash[b]{(N+1)(N+2)}}}\,\dww-\,\ws\,\tfrac{1}{\raisebox{.09in}~[2(N+1)]^{3/2}}-\,\ws\,\tfrac{1}{4(N+1)\sqrt{\smash[b]{2(N+2)}}}\,\dww~\Big]\,,\label{Xmu}\\[5pt]
		\hspace{-.1cm}	\mathrm{Y}~&:=-\,\ds\,+\,(\w\c\p)\,\wbs\,+\,\ws\,\ds\,\tfrac{1}{\sqrt{\smash[b]{2(N+1)}}}\,-\,\ws\,(\w\c\p)\,\tfrac{1}{\sqrt{\smash[b]{2(N+2)}}}\,\wbs \,, \label{Y}\\[5pt]
		\hspace{-.1cm}	\mathrm{Y}_\m&:= \m\,\Big[~\tfrac{1}{N+1}\,-\,\ws\,\tfrac{1}{2(N+2)}\,\wbs\,+\,\tfrac{1}{\sqrt{\smash[b]{2(N+1)}}}\,\wbs\,-\,\w^2\,\tfrac{1}{2(N+1)\,\sqrt{\smash[b]{2(N+2)}}}\,\wbs\,-\,\w^2\,\tfrac{1}{2(N+1)\,\sqrt{\smash[b]{(N+1)(N+2)}}}   ~\Big] \,,\label{Ymu}\\[5pt]
		\hspace{-.1cm}	\mathrm{Z}~&:=1\,-\,\ws\,\tfrac{1}{2(N+1)}\,\wbs\,+\,\tfrac{1}{\sqrt{\smash[b]{2(N+1)}}}\,\wbs \,.\label{Z}
	\end{align}
	%
	%
	We note that the operators $\mathrm{T}$, $\mathrm{X}$, $\mathrm{X}_\m$ were obtained in \cite{Najafizadeh:2019mun} for an on-shell description, while the operators $\mathrm{Y}$, $\mathrm{Y}_\m$, $\mathrm{Z}$ were found in this work to make an off-shell description.
	
	\vspace{.5cm}
	
	Let us again check the closure of the supersymmetry algebra using the above supersymmetry transformations \eqref{susy tr cons}. In comparison to unconstrained formulation, here, a long but straightforward computation implies that the commutator of the SUSY transformations \eqref{susy tr cons} on the CSP gauge fields yield
	\begin{subequations}  
		\label{closur con}	
		\begin{align}
			[\,\d_1\,,\,\d_2\,]\,\Phi(x,\w)&=\,2\,(\bar\ep\2\,\ds\,\ep\1)\,\Phi(x,\w)\,,\label{b clou1 c}\\[5pt]
			[\,\d_1\,,\,\d_2\,]\,\mathrm{H}(x,\w)&=\,2\,(\bar\ep\2\,\ds\,\ep\1)\,\mathrm{H}(x,\w)\,,\label{b clou2 c}\\[5pt]
			[\,\d_1\,,\,\d_2\,]\,\Psi(x,\w)&=\,2\,(\bar\ep\2\,\ds\,\ep\1)\,\Psi(x,\w)\,+\,\hbox{{\small G.T.}}\,,\label{f clou c}
		\end{align}
	\end{subequations}  
	where $\hbox{{\small G.T.}}$ denotes a field-dependent spinor gauge transformation corresponding to \eqref{Gauge T F 2}. Thus, the algebra closes off-shell, up to a gauge transformation, that is, no equation of motion has been applied.
	
	\vspace{.5cm}
	
	Let us now reproduce on-shell supersymmetry transformations \cite{Najafizadeh:2019mun} by applying the equations of motion on the off-shell supersymmetry transformations \eqref{susy tr cons}. For this purpose, using the following identity\,\footnote{Due to the triple gamma-trace condition \eqref{trace conditions f}, terms containing of $\omega\hspace{-7pt}\slash\hspace{3pt}^3$ and (or) $\overbar\omega\hspace{-7pt}\slash\hspace{3pt}^3$ vanish at the level of the action, so such terms do not contribute to the supersymmetry transformations \eqref{susy tr cons}.}
	\begin{align}
		\big[\,\mathrm{Y} \,+\, \mathrm{Y}_\m\,\big]\,\Psi(x,\w)=	\big(\,1\,-\,\ws\,\tfrac{1}{~[\,2(N+1)\,]^{1/2}}\,\big)\,\big(\,\mathcal{F} \,+\, \mathcal{F}_\m\,\big)\,\Psi(x,\w)\,,
	\end{align}
	one may rewrite \eqref{susy tr cons2} to emerge explicitly the Dirac continuous spin equation of motion \eqref{F+Fmu--}. By this rewriting, one can then apply equations of motion \eqref{eom H-},\eqref{F+Fmu--} in the off-shell supersymmetry transformations \eqref{susy tr cons}, resulting in the on-shell supersymmetry transformations (obtained in \cite{Najafizadeh:2019mun})
	\begin{align}
		\d \,\Phi &={\scriptstyle\sqrt{2}}\,\,\bar\ep~L~\mathrm{T}~{\Psi}\,, \qquad\qquad  \d \,\Psi ={\scriptstyle\sqrt{2}} \,\,\big[\, \mathrm{X} \,+\, \mathrm{X}_\m \,\big]\, R~\ep~\Phi\,,
	\end{align}
	where $L$, $R$, $T$, $X$, $X_\m$ were respectively introduced in \eqref{projections},\eqref{T},\eqref{X},\eqref{Xmu}.
	
	\section{Reformulation in terms of real fields} \label{reformulation}
	
	In both unconstrained and constrained Lagrangian formulations \ref{unconstrained formu},\ref{constrained formu}, we presented supersymmetry actions, and their associated supersymmetry transformations, in terms of complex fields $\Phi$, $\mathrm{H}$, $\Psi$. The great advantage of this consideration is that the SUSY actions \eqref{off susy u action},\eqref{susy off const} and transformations \eqref{susy tr},\eqref{susy tr cons} have taken a simple compact form, and hence less calculations are required to see action invariance and check the closure of the SUSY algebra. However, as one can see, in terms of real fields they would take a more complicated form. Nevertheless, reformulation in terms of real fields might be beneficial in order to reproduce higher spin results. This approach has been done in unconstrained formulation for on-shell description with $\m=0$ \cite{Najafizadeh:2020moz}, and one can conveniently extend it to the off-shell description with $\m\neq0$. Therefore, leaving unconstrained formalism, here, we just discuss constrained formalism \ref{constrained formu} and reformulate it in terms of real fields. This reformulation will be used in the next section, where we shall extract higher spin results.
	
	\subsection{Constrained formalism}
	
	Let us consider the complex scalar $\Phi$, auxiliary complex scalar $\mathrm{H}$, and Dirac $\Psi$ CSP gauge fields as 
	\begin{subequations}
		\label{real}
		\begin{align} 
			\Phi(x,\w)&=\tfrac{1}{\sqrt{2}}\,\le[\,\phi(x,\w)\,-\,i\,\phi'(x,\w)\,\ri]\,,\label{real phi}\\[5pt] 
			\mathrm{H}(x,\w)&=\tfrac{1}{\sqrt{2}}\,\le[\,h(x,\w)\,+\,i\,h'(x,\w)\,\ri]\,,\label{real h}\\[5pt]
			\Psi(x,\w)&=\tfrac{1}{\sqrt{2}}\,\le[\,\psi(x,\w)\,-\,i\,\psi'(x,\w)\,\ri]\,,\label{real psi}
		\end{align} 
	\end{subequations}
	where $\phi$, $\phi'$ are real scalar continuous spin gauge fields; $h$, $h'$ are auxiliary real scalar continuous spin gauge fields\,\footnote{One may refer to $\phi$, $\phi'$ (and thus $h$, $h'$) as two bosonic fields which have opposite parities.}; and $\psi$, $\psi'$ are two Majorana continuous spin gauge fields. By this consideration, the off-shell supersymmetric continuous spin action \eqref{susy off const} reads
	\begin{align} 
		\hspace{0cm}S_{_{\hbox{{\tiny \,SUSY}}}}^{^{\hbox{{\,\tiny CSP}}}}&={ \tfrac{1}{2}\int d^4 x\, \Big[\,\phi(x,\wb)\,\big(\,\mathrm{B}+\mathrm{B}_\m\,\big)\,{\phi}(x,\w)\,+\,h(x,\wb)\,\big(\,{\mathrm{B}_{_0}}\,\big)\,h(x,\w)\,+\,\overline{\psi}(x,\wb)\,\big(\,\mathrm{F}+\mathrm{F}_\m\,\big)\,\psi(x,\w)}
		\label{real csp action}\\[4pt]
		\hspace{-.4cm}&\qquad~~~+\,{ \phi'(x,\wb)\,\big(\,\mathrm{B}+\mathrm{B}_\m\,\big)\,\phi'(x,\w)\,+\,h'(x,\wb)\,\big(\,{\mathrm{B}_{_0}}\,\big)\,h'(x,\w)\,+\,\overline{\psi'}(x,\wb)\,\big(\,\mathrm{F}+\mathrm{F}_\m\,\big)\,\psi'(x,\w)
			\Big]\Big|_{\w=0}},\nonumber
	\end{align}
	where $\mathrm{B}$, $\mathrm{B}_\m$, $\mathrm{B}_{_0}$, $\mathrm{F}$, $\mathrm{F}_\m$ were introduced in \eqref{susy off const}. To avoid repetition, let us just notice that the real continuous spin gauge fields $\phi$, $h$, $\psi$ (and $\phi'$, $h'$, $\psi'$) have respectively similar gauge symmetries as \eqref{Gauge T m},\eqref{GT H},\eqref{Gauge T F 2}, introduced by equivalent generating functions as \eqref{Phi, xi,w},\eqref{generating func H},\eqref{Psi, tau,w}, and satisfy analogous constraints as \eqref{trace conditions},\eqref{trace cond H},\eqref{trace conditions f}.
	
	\vspace{.5cm}
	
	Therefore, we find that the off-shell supersymmetric continuous spin action \eqref{real csp action} is invariant under the following supersymmetry transformations 
	\begin{subequations}
		\label{real susy tr}  
		\begin{align}
			\hspace{-.2cm}	\d\phi{\,}&=\tfrac{1}{\sqrt{2}}\,\bar\epsilon\,\big[\,\mathrm{T}\,\psi\,\,-\,i\g^5\,\mathrm{T}\,\psi'\, \big]\,,\qquad\qquad\qquad\qquad\qquad\,\,
			\d\phi'=\tfrac{1}{\sqrt{2}}\,\bar\epsilon\,\big[\,\mathrm{T}\,\psi'\,+\,i\g^5\,\mathrm{T}\,\psi\, \big]\,,\\[16pt]
			\hspace{-.2cm}	\d h &=\tfrac{1}{\sqrt{2}}\,\bar\epsilon\,\big[(\mathrm{Y}+\mathrm{Y}_\m)\,\psi\,\,+\,i\g^5\,(\mathrm{Y}+\mathrm{Y}_\m)\,
			\psi'\, \big]\,,\qquad~~~\,\,\,\,\,
			\d h'=\tfrac{1}{\sqrt{2}}\,\bar\epsilon\,\big[\!-(\mathrm{Y}+\mathrm{Y}_\m)\,\psi'\,+\,i\g^5\,(\mathrm{Y}+\mathrm{Y}_\m)\,
			\psi\,\, \big]\,,\\[3pt]
			\hspace{-.2cm}	\d\psi&=\tfrac{1}{\sqrt{2}}\le\{(\mathrm{X}+\mathrm{X}_\m)\big[\phi+i\g^5\phi' \big]-\mathrm{Z}\,\big[h+i\g^5h' \big]\ri\}\epsilon\,,\quad
			\d\psi'=\tfrac{1}{\sqrt{2}}\le\{(\mathrm{X}+\mathrm{X}_\m)\big[\phi'-i\g^5\phi \big]+\mathrm{Z}\,\big[h'-i\g^5 h \big]\ri\}\epsilon\,,
		\end{align}
	\end{subequations}
	where $\epsilon$ is now an arbitrary constant Majorana spinor, and $\mathrm{T}$, $\mathrm{X}$, $\mathrm{X}_\m$, $\mathrm{Y}$, $\mathrm{Y}_\m$, $\mathrm{Z}$ were given in \eqref{T}-\eqref{Z}. As we mentioned, although reformulation in terms of real fields becomes sophisticated, it allows us to recover higher spin results in a more convenient way.   
	
	
	\section{Reducing to the higher spin supermultiplets} \label{Extraction}
	
	Let us recall again that when continuous spin parameter $\m$ vanishes, the continuous spin representation becomes reducible and decomposes into a direct sum of all helicity representations. Therefore, in this section, we expect in the $\m\rightarrow 0$ limit the off-shell supersymmetric continuous spin action \eqref{real csp action} and its supersymmetry transformations \eqref{real susy tr} reduce to off-shell actions and supersymmetry transformations of the scalar supermultiplet $(\,{\scriptstyle 0}\,,\, {\scriptstyle 1/2}\,)$; integer spin supermultiplets $(\,s\,,\, s\, {\scriptstyle +\, 1/2}\,),\, s\geqslant1$; and half-integer spin supermultiplets $(\, s\, {\scriptstyle -\, 1/2}\,,\, s \,\,),\, s\geqslant1$. Here is a step by step explanation of how to obtain these results.

	\subsection{Off-shell supersymmetric higher spin theory}
	At $\m=0$, the off-shell SUSY CSP action \eqref{real csp action} reduces to the so called ``off-shell supersymmetric higher spin action''
	\begin{align} 
		S_{_{\hbox{{\tiny \,SUSY}}}}^{^{\hbox{{\,\tiny HS}}}}&=\tfrac{1}{2}\int d^4 x\, \Big[\,\phi(x,\wb)\,\big(\,\mathrm{B}\,\big)\,{\phi}(x,\w)\,+\,h(x,\wb)\,\big(\,{\mathrm{B}_{_0}}\,\big)\,h(x,\w)\,+\,\overline{\psi}(x,\wb)\,\big(\,\mathrm{F}\,\big)\,\psi(x,\w)
		\nonumber\\[0pt]
		&\qquad~~\,~~~\,+\,\phi'(x,\wb)\,\big(\,\mathrm{B}\,\big)\,\phi'(x,\w)\,+\,h'(x,\wb)\,\big(\,{\mathrm{B}_{_0}}\,\big)\,h'(x,\w)\,+\,\overline{\psi'}(x,\wb)\,\big(\,\mathrm{F}\,\big)\,\psi'(x,\w)
		\Big]\Big|_{\w=0},\label{real hs action}
	\end{align}
	where the operators $\mathrm{B}$, $\mathrm{B}_{_0}$, $\mathrm{F}$ were given in \eqref{susy off const}. We note that the bosonic $\mathrm{B}$ and the fermionic $\mathrm{F}$ operators are respectively related to the Fronsdal and Fang-Fronsdal operators. More precisely, in the above action, the terms including of $\mathrm{B}$ describe a direct sum of all Fronsdal actions \cite{Fronsdal:1978rb}, and the expressions containing of $\mathrm{F}$ characterize a direct sum of all Fang-Fronsdal actions \cite{Fang:1978wz}. Since we set $\m=0$, the entered real fields $\phi$, $h$, $\psi$ (and $\phi'$, $h'$, $\psi'$) in the action \eqref{real hs action} are now higher spin gauge fields, introduced by the generating functions\,\footnote{In the generating functions, $\phi_{\m_1 \dots \m_s}$ denotes a collection of totally symmetric real tensor fields of all integer rank $s$, $h_{\m_1 \dots \m_s}$ stands for a collection of totally symmetric auxiliary real tensor fields of all integer rank $s$, and $\psi_{\m_1 \dots \m_s}$ indicates a set of totally symmetric Majorana spinor-tensor fields of all half-integer rank $s+{\scriptstyle 1/2}$}
	\begin{align}
		\phi(x,\w)&{\small =\suz~\phi_s(x,\w)\,, \qquad\qquad~~~ h(x,\w)=\suz~ h_s(x,\w)\,,~\,\qquad\quad~\,~ \psi(x,\w)=\suz~\psi_s(x,\w)\,,\label{gener HS}}\\[4pt]
		&{\small =\suz\tfrac{1}{s!}\,\w^{\m_1}\!\dots \w^{\m_s}\phi_{\m_1 \dots \m_s}\,,\qquad~~~\,
			=\suz\tfrac{1}{s!}\,\w^{\m_1}\!\dots \w^{\m_s}h_{\m_1 \dots \m_s}\,, \qquad ~~\,
			=\suz\tfrac{1}{s!}\,\w^{\m_1}\!\dots \w^{\m_s}\psi_{\m_1 \dots \m_s}}\,.\nonumber
	\end{align} 
	Here, the gauge symmetry of the real higher spin gauge fields $\phi$, $h$, $\psi$ (and similarly $\phi'$, $h'$, $\psi'$) read
	\be 
	\d\,\phi(x,\w)=(\w\c\p)\,\chi(x,\w)\,,\qquad\qquad \d\,h(x,\w)=0\,, \qquad\qquad \d\,\psi(x,\w)=(\w\c\p)\,\tau(x,\w)\,,
	\ee 
	where $\phi, h$ are double-traceless, $\chi$ is traceless, and $\psi, \tau$ are triple gamma-traceless and gamma-traceless.
	
	
	\vspace{.5cm}
	
	Now, by setting $\m=0$ in the SUSY transformations \eqref{real susy tr}, one then finds that the off-shell supersymmetric higher spin action \eqref{real hs action} is invariant under the following supersymmetry transformations
	\begin{subequations}
		\label{real susy tr hs}  
		\begin{align}
			\d\phi{\,}&=\tfrac{1}{\sqrt{2}}\,\bar\epsilon\,\big[\,\mathrm{T}\,\psi\,\,-\,i\g^5\,\mathrm{T}\,\psi'\, \big]\,,\qquad\qquad\qquad\,\,~\,\qquad\quad
			\d\phi'=\tfrac{1}{\sqrt{2}}\,\bar\epsilon\,\big[\,\mathrm{T}\,\psi'\,+\,i\g^5\,\mathrm{T}\,\psi\, \big]\,,\\[4pt]
			\d h &=\tfrac{1}{\sqrt{2}}\,\bar\epsilon\,\big[\mathrm{Y}\,\psi\,\,+\,i\g^5\,\mathrm{Y}\,
			\psi'\, \big]\,,\qquad~~~\quad\quad\quad\qquad\quad\,\,\,\,\,
			\d h'=\tfrac{1}{\sqrt{2}}\,\bar\epsilon\,\big[\!-\mathrm{Y}\,\psi'\,+\,i\g^5\,\mathrm{Y}\,\psi\,\, \big]\,,\\[4pt]
			\d\psi&=\tfrac{1}{\sqrt{2}}\le\{\mathrm{X}\,\big[\phi+i\g^5\phi' \big]-\mathrm{Z}\,\big[h+i\g^5h' \big]\ri\}\epsilon\,,\quad\quad\quad~\,\,
			\d\psi'=\tfrac{1}{\sqrt{2}}\le\{\mathrm{X}\,\big[\phi'-i\g^5\phi \big]+\mathrm{Z}\,\big[h'-i\g^5 h \big]\ri\}\epsilon\,,
		\end{align}
	\end{subequations}
	where the operators $\mathrm{T}$, $\mathrm{X}$, $\mathrm{Y}$, $\mathrm{Z}$ were introduced in \eqref{T},\eqref{X},\eqref{Y},\eqref{Z} respectively. 
	
	\vspace{.5cm}
	
	We recall that in the on-shell 4d, $\mathcal{N}=1$ massless higher spin supermultiplets \`a la Fronsdal \cite{Curtright:1979uz}, the spin-$0$ field was not considered in the spectrum\,\footnote{Also in off-shell superfield description \cite{Kuzenko:1993jq, Kuzenko:1993jp} and its extension to AdS \cite{Kuzenko:1994dm}. However, the spin-$0$ field is included in 4-dimensional flat \cite{Najafizadeh:2019mun} and AdS spaces \cite{Gates:1996xs, Najafizadeh:2020moz}.}. Nevertheless, if one wants to include the spin-$0$ field, one should take into account the off/on-shell supersymmetric higher spin action and its supersymmetry transformations as \eqref{real hs action},\eqref{real susy tr hs} respectively. Note that we arrived to this result from a top-down approach (i.e. from the CSP gauge theory), in which the spin-$0$ field has already existed in the spectrum. Hence, in order to separate the spin-$0$ field, let us first decompose the action \eqref{real hs action}, and then proceed further.

	\vspace{.5cm}
	
	\noindent\textbf{Action decomposition}
	
	\vspace{.3cm}
	
	\noindent The off-shell SUSY higher spin action \eqref{real hs action} is comprising of the actions of the scalar supermultiplet, integer spin supermultiplets, and half-integer spin supermultiplets which are mixed together. To determine the off-shell actions corresponding to each supermultiplet, by considering \eqref{gener HS}, let us separate the first term in the generating functions of $\phi$, $\phi'$, $h$, $h'$, $\psi$, except $\psi'$. For example consider
	\be
	\phi(x,\w)=\phi(x)+\suo~\phi_s(x,\w)\,.
	\ee
	By this separation, not applying to $\psi'$, we can rewrite the off-shell SUSY higher spin action \eqref{real hs action} as
	\begin{subequations}
		\label{3 action}
		\be 
		S_{_{\hbox{{\tiny \,SUSY}}}}^{^{\hbox{{\,\tiny HS}}}}~=~
		S_{_{\hbox{{\tiny $(\,{\scriptstyle 0} , {\scriptstyle 1/2}\,)$ }}}}^{^{\hbox{{\,\tiny }}}}~+~
		\suo~S_{_{\hbox{{\tiny $(s, s \,{\scriptstyle + 1/2})$}}}}^{^{\hbox{{\,\tiny }}}}~+~
		\suo~S_{_{\hbox{{\tiny $(s \,{\scriptstyle - 1/2}, s)$}}}}^{^{\hbox{{\,\tiny }}}}\,,
		\ee  
		in which 
		\begin{align}
			S_{_{\hbox{{\tiny $(\,{\scriptstyle 0} , {\scriptstyle 1/2}\,)$}}}}&=
			\tfrac{1}{2}\,\int d^4 x\,\Big[\,\phi(x)\,\Box\,\phi(x)\,+\,\phi'(x)\,\Box\,\phi'(x)\,+\,[h(x)]^2\,+\,[h'(x)]^2\,-\,\overbar{\psi}(x)\,\ds\,\psi(x)\,\Big]\,, \label{scalar sup}
			\\[9pt]
			S_{_{\hbox{{\tiny $(s, s \,{\scriptstyle + 1/2})$}}}}&=
			\tfrac{1}{2}\,\int d^4 x\, \Big[\,\phi_s(x,\wb)\,\big(\,\mathrm{B}\,\big)\,{\phi}_s(x,\w)\,+\,h_s(x,\wb)\,\big(\,{\mathrm{B}_{_0}}\,\big)\,h_s(x,\w)\,+\,\overline{\psi}_s(x,\wb)\,\big(\,\mathrm{F}\,\big)\,\psi_s(x,\w)\,\Big]\Big|_{\w=0}\,, \label{integer sup}
			\\[4pt] 
			S_{_{\hbox{{\tiny $(s \,{\scriptstyle - 1/2}, s)$}}}}&=
			\tfrac{1}{2}\,\int d^4 x\, \Big[\,\phi'_s(x,\wb)\,\big(\,\mathrm{B}\,\big)\,\phi'_s(x,\w)\,+\,h'_s(x,\wb)\,\big(\,{\mathrm{B}_{_0}}\,\big)\,h_s'(x,\w)\,+\,\overline{\psi'}_{s-1}(x,\wb)\,\big(\,\mathrm{F}\,\big)\,\psi_{s-1}'(x,\w)
			\Big]\Big|_{\w=0}\,,\label{half-integer sup}
		\end{align}
	\end{subequations}
	are respectively the scalar supermultiplet action, the integer spin supermultiplet action, and the half-integer spin supermultiplet action. Concerning the scalar supermultiplet action \eqref{scalar sup}, we note that the first term in the generating functions has no dependence to $\w$. Therefore, in the action \eqref{scalar sup}, the bosonic $\mathrm{B}$, $\mathrm{B}_{_0}$ and the fermionic $\mathrm{F}$ operators acting on these fields have been reduced to: $\mathrm{B}=\Box$, $\mathrm{B}_{_0}=1$, $\mathrm{F}=-\,\ds$ (see the form of these operators in \eqref{susy off const}).   
	
	\vspace{.5cm}
	
	Up to now, we decomposed the off-shell SUSY higher spin action \eqref{real hs action} into the well-known supermultiplet actions \eqref{3 action}. The next duty is reading supersymmetry transformations corresponding to each supermultiplet from the higher spin SUSY transformations \eqref{real susy tr hs}. To this end, using \eqref{T},\eqref{X},\eqref{Y},\eqref{Z}, we divide each entered operator in the supersymmetry transformations \eqref{real susy tr hs} into two parts 
	\be
	\mathrm{T}=\mathrm{T}_1+\mathrm{T}_2\,,\qquad\qquad
	\mathrm{X}=\mathrm{X}_1+\mathrm{X}_2\,,\qquad\qquad
	\mathrm{Y}=\mathrm{Y}_1+\mathrm{Y}_2\,,\qquad\qquad
	\mathrm{Z}=\mathrm{Z}_1+\mathrm{Z}_2\,,\label{operators decom}
	\ee 
	where
	\begin{subequations}
		\label{index 1}
		\begin{align}
			\mathrm{T}\1&:=1\,,   \\[5pt]
			\mathrm{X}\1&:=\,\ds-\ws\tfrac{1}{2(N+1)}(\dw\c\p)+\ws\ds\tfrac{1}{2(N+1)}\wbs - \ws(\w\c\p)\tfrac{1}{4(N+2)}\dww \,,\\[5pt]
			\mathrm{Y}\1&:=-\,\ds\,+\,(\w\c\p)\,\wbs~=~\mathcal{F}\,,  \\[5pt]
			\mathrm{Z}\1&:=1\,-\,\ws\,\tfrac{1}{2(N+1)}\,\wbs\,, 
		\end{align}	
	\end{subequations}
	and
	\begin{subequations}
		\label{index 2}
		\begin{align}			
			\mathrm{T}\2&:=\ws\,\tfrac{1}{\sqrt{\smash[b]{2(N+1)}}}\,,\\[5pt]
			\mathrm{X}\2&:=\tfrac{1}{\sqrt{\smash[b]{2(N+1)}}}\big[\dwdx-\ds\,\wbs+ \tfrac{1}{2}\wdx\dww\big]\,,\\[5pt]
			\mathrm{Y}\2&:=\ws\,\ds\,\tfrac{1}{\sqrt{\smash[b]{2(N+1)}}}-\,\ws\,(\w\c\p)\,\tfrac{1}{\sqrt{\smash[b]{2(N+2)}}}\,\wbs~=~\big[\,\ws\,\tfrac{-\,1}{\sqrt{\smash[b]{2(N+1)}}}\,\big]\,\mathcal{F}\,,\\[5pt]
			\mathrm{Z}\2&:=\tfrac{1}{\sqrt{\smash[b]{2(N+1)}}}\,\wbs \,.
		\end{align} 
	\end{subequations} 
	Note that the operators $\mathrm{Y}\1$, $\mathrm{Y}\2$ are both proportional to the Fang-Fronsdal operator $\mathcal{F}$ \eqref{F-F operator}. This decomposition was done so as operators with index 1 have no degree of $\w$ or $\wb$, while operators with index 2 have a degree of $\w$ or $\wb$ (the degree of $\w$ is $1$, thus the degree of $\wb\,(:=\p/\p \w)$ would be $-1$). Indeed, we will see below that each set of operators \eqref{index 1},\eqref{index 2} will appear in a specific supermultiplet. Referring to \eqref{real susy tr hs},\eqref{3 action}, let us read related SUSY transformations of each supermultiplet in below.

	\subsection{Scalar (chiral) supermultiplet: $(\,{\scriptstyle 0}\,,\, {\scriptstyle 1/2}\,)$}
	The first part of \eqref{3 action} describes the off-shell scalar supermultiplet action (off-shell Wess-Zumino model)
	\be 
	S_{_{\hbox{{\tiny $(\,{\scriptstyle 0} , {\scriptstyle 1/2}\,)$}}}}=
	\tfrac{1}{2}\,\int d^4 x\,\Big[~\,\phi\,\Box\,\phi\,+\,\phi'\,\Box\,\phi'\,+\,h^2\,+\,h'^{\,2}\,-\,\overbar{\psi}\,\ds\,\psi\,~\Big]\,, \label{wz action}
	\ee 
	in which $\phi$ (or $\phi'$) is the spin-$0$ field, $h$ (or $h'$) is an auxiliary spin-$0$ field, and $\psi$ is the spin-${\scriptstyle {1}/{2}}$ field (Majorana spinor). Since the gauge field $\psi'$ does not contribute to this action \eqref{wz action}, one may set $\psi'=0$ in the higher spin SUSY transformations \eqref{real susy tr hs}. On the other side, gauge fields in \eqref{wz action} are the first term of the generating functions having no dependence to $\w$. This demonstrates that entered operators in \eqref{real susy tr hs}, acting on the gauge fields, result in
	\be
	\mathrm{T}=\mathrm{T}\1=1\,,\qquad \mathrm{X}=\mathrm{X}\1=\ds\,, \qquad \mathrm{Y}=\mathrm{Y}\1=-\,\ds\,,\qquad\mathrm{Z}=\mathrm{Z}\1=1\,.\label{wz operators}
	\ee   
	Therefore, $(i)$ by throwing out the fermionic field $\psi'$ and considering the first term of generating functions, $(ii)$ applying \eqref{wz operators}, and $(iii)$ redefining supersymmetry parameter $\ep:=\epsilon{\scriptstyle /\sqrt{2}}$, the off-shell higher spin SUSY transformations \eqref{real susy tr hs} reduce to 	  
	\begin{align}
		\d\phi{\,}&=\bar{\ep}\,\psi\,,\qquad\qquad\quad~~~~\,
		\d h{\,} =-\,\bar\ep\,\ds\,\psi\,,\qquad\qquad\quad \d\psi=\ds\,\big[\,\phi+i\g^5\phi' \,\big]\,\ep\,-\,\big[\,h+i\g^5h' \,\big]\,\ep\,,\label{wz susy tra}\\[4pt]
		\d\phi'&=\bar\ep\,i\g^5\psi	 \,,\qquad\qquad\quad
		\d h'=- \,\bar\ep\,i\g^5\,\ds\,\psi \,.\nonumber
	\end{align}
	These are precisely SUSY transformations of the scalar supermultiplet $(\,{\scriptstyle 0}\,,\, {\scriptstyle 1/2}\,)$, leaving the action \eqref{wz action} invariant. Let us emphasize that this result was obtained from the off-shell continuous spin supermultiplet in a limit (see Fig.\,\ref{fig3}).

	\subsection{Integer spin supermultiplets: $(\,s\,,\, s\, {\scriptstyle +\, 1/2}\,),\, s\geqslant1$}
	The off-shell integer spin supermultiplet action can be given by (see \cite{Najafizadeh:2019mun} for a review on on-shell description)
	\be 
	S_{_{\hbox{{\tiny $(s, s \,{\scriptstyle + 1/2})$}}}}=
	\tfrac{1}{2}\,\int d^4 x\, \Big[\,\phi_s(x,\wb)\,\big(\,\mathrm{B}\,\big)\,{\phi}_s(x,\w)\,+\,h_s(x,\wb)\,\big(\,{\mathrm{B}_{_0}}\,\big)\,h_s(x,\w)\,+\,\overline{\psi}_s(x,\wb)\,\big(\,\mathrm{F}\,\big)\,\psi_s(x,\w)\,\Big]\Big|_{\w=0}\,, \label{integer sup-}
	\ee
	which is the second part of the SUSY higher spin action \eqref{3 action}. The gauge fields $\phi_s$, $h_s$, $\psi_s$ are given by generating functions \eqref{gener HS}, and operators $\mathrm{B}$, $\mathrm{B}_{_0}$, $\mathrm{F}$ were defined in \eqref{susy off const}. We notice that, the first and the last parts of the action \eqref{integer sup-} are precisely the Fronsdal \cite{Fronsdal:1978rb} and the Fang-Fronsdal \cite{Fang:1978wz} actions respectively. Since the action \eqref{integer sup-} does not consist of $\phi'$, $h'$, $\psi'$, one may thus drop these fields in the SUSY transformations \eqref{real susy tr hs}. On the other side, generating functions in the integer spin supermultiplet ($\phi_s$, $h_s$\,;\, $\psi_s$) have equal number of $\w$, thus entered operators in SUSY transformations should not have a degree of $\w$ or $\wb$. In other words, entered operators in \eqref{real susy tr hs} should be considered with index 1 \eqref{index 1}, i.e.
	\be
	\mathrm{T}=\mathrm{T}\1\,,\qquad \mathrm{X}=\mathrm{X}\1\,, \qquad \mathrm{Y}=\mathrm{Y}\1\,,\qquad\mathrm{Z}=\mathrm{Z}\1\,.\label{operators 1}
	\ee    
	Therefore, $(i)$ by dropping $\phi'$, $h'$, $\psi'$ and keeping $\phi_s$, $h_s$, $\psi_s$, $(ii)$ applying \eqref{operators 1}, and $(iii)$ redefining supersymmetry parameter $\ep:=\epsilon{\scriptstyle /2}$, the off-shell higher spin SUSY transformations \eqref{real susy tr hs} turn into 	  
	\begin{subequations}
		\label{s,s+1/2} 
		\begin{align}
			\d \,\phi_s &={\scriptstyle\sqrt{2}}\,\,\bar\ep~\mathrm{T}\1\,{\psi_s}\,, \\[5pt]
			\d \,h_s&= {\scriptstyle\sqrt{2}}\,\,\bar\ep~\mathrm{Y}\1\,\psi_s\,,  \\[5pt]
			\d \,\psi_s &={\scriptstyle\sqrt{2}} ~\mathrm{X}\1\,\ep\,\phi_s\,-\,
			{\scriptstyle\sqrt{2}} ~\mathrm{Z}\1\,\ep\,h_s\,,
		\end{align}
	\end{subequations}
	where $\mathrm{T}\1$, $\mathrm{X}\1$, $\mathrm{Y}\1$, $\mathrm{Z}\1$ introduced in \eqref{index 1}. These are indeed off-shell SUSY transformations of the integer spin supermultiplet, leaving the action \eqref{integer sup-} invariant. It is easy to see that obtained SUSY transformations \eqref{s,s+1/2} recover on-shell results \cite{Curtright:1979uz} (see \cite{Najafizadeh:2019mun} for more detail)
	\begin{align}
		\d \,\phi_s &={\scriptstyle\sqrt{2}}\,\,\bar\ep~\mathrm{T}\1\,{\psi_s}\,, \qquad\qquad
		\d \,\psi_s ={\scriptstyle\sqrt{2}} ~\mathrm{X}\1\,\ep\,\phi_s\,, \label{s,s+1/2 on}
	\end{align}
	if one applies the fermionic equation of motion $\mathrm{Y}\1\,\psi_s=\mathcal{F}\,\psi_s=0$, and the auxiliary equation $h_s=0$.

	\subsection{Half-integer spin supermultiplets: $(\, s\, {\scriptstyle -\, 1/2}\,,\, s \,\,),\, s\geqslant1$}
	The off-shell half-integer spin supermultiplet action can be given by 
	\be 
	S_{_{\hbox{{\tiny $(s \,{\scriptstyle - 1/2}, s)$}}}}=
	\tfrac{1}{2}\int d^4 x\, \Big[\,\phi'_s(x,\wb)\,\big(\,\mathrm{B}\,\big)\,\phi'_s(x,\w)\,+\,h'_s(x,\wb)\,\big(\,{\mathrm{B}_{_0}}\,\big)\,h_s'(x,\w)\,+\,\overline{\psi'}_{s-1}(x,\wb)\,\big(\,\mathrm{F}\,\big)\,\psi_{s-1}'(x,\w)
	\Big]\Big|_{\w=0},\label{half-integer sup-}
	\ee
	which is the last part of the SUSY higher spin action \eqref{3 action}. The contributing fields $\phi'_s$, $h'_s$, $\psi'_{s-1}$ are generating functions \eqref{gener HS}, and operators $\mathrm{B}$, $\mathrm{B}_{_0}$, $\mathrm{F}$ were defined in \eqref{susy off const}. Since the action \eqref{half-integer sup-} does not include $\phi$, $h$, $\psi$, one may throw out these fields in the SUSY transformations \eqref{real susy tr hs}. On the other hand, generating functions in the half-integer spin supermultiplet ($\phi'_s$, $h'_s$\,;\, $\psi'_{s-1}$) does not have equal number of $\w$ for an arbitrary spin ($s\geqslant1$). Thus, entered operators in SUSY transformations should have a degree of $\w$ or $\wb$ to compensate this inequality. In other words, the entered operators in \eqref{real susy tr hs} should be taken into account with index 2 \eqref{index 2}, i.e.
	\be
	\mathrm{T}=\mathrm{T}\2\,,\qquad \mathrm{X}=\mathrm{X}\2\,, \qquad \mathrm{Y}=\mathrm{Y}\2\,,\qquad\mathrm{Z}=\mathrm{Z}\2\,.\label{operators 2}
	\ee    
	Therefore, $(i)$ by omitting $\phi$, $h$, $\psi$ and preserving $\phi'_s$, $h'_s$, $\psi'_{s-1}$, $(ii)$ applying \eqref{operators 2}, and $(iii)$ redefining supersymmetry parameter $\ep:=\epsilon{\scriptstyle /2}$, the off-shell higher spin SUSY transformations \eqref{real susy tr hs} become  
	\begin{subequations}
		\label{s-1/2,s} 
		\begin{align}
			\d \,\phi'_s =~&{\scriptstyle\sqrt{2}}\,\,\bar\ep~\mathrm{T}\2\,{\psi'_{s-1}}\,, \label{susy tr cons11}\\[5pt]
			\d \,h'_s=-\, &{\scriptstyle\sqrt{2}}\,\,\bar\ep~\mathrm{Y}\2\,\psi'_{s-1}\,,\label{susy tr cons22}  \\[5pt]
			\d \,\psi'_{s-1} =~&{\scriptstyle\sqrt{2}} ~\mathrm{X}\2\,\ep\,\phi'_s\,+\,
			{\scriptstyle\sqrt{2}} ~\mathrm{Z}\2\,\ep\,h'_s\,,\label{susy tr cons33}
		\end{align}
	\end{subequations}
	where $\mathrm{T}\2$, $\mathrm{X}\2$, $\mathrm{Y}\2$, $\mathrm{Z}\2$ introduced in \eqref{index 2}. These are off-shell SUSY transformations of the half-integer spin supermultiplet, leaving the action \eqref{half-integer sup-} invariant. In particular, for $s=2$, the action \eqref{half-integer sup-} and SUSY transformations \eqref{s-1/2,s} describe the off-shell 4d $\mathcal{N}=1$ linearized supergravity multiplet in flat spacetime. Let us again bring our attention to the fact that these results were discovered from the continuous spin supermultiplet in a limit (see Fig.\,\ref{fig3}). The acquired SUSY transformations \eqref{s-1/2,s} will recover on-shell results \cite{Curtright:1979uz} (see \cite{Najafizadeh:2019mun} for more detail)
	\begin{align}
		\d \,\phi'_s =~&{\scriptstyle\sqrt{2}}\,\,\bar\ep~\mathrm{T}\2\,{\psi'_{s-1}}\,, \qquad\qquad
		\d \,\psi'_{s-1} =~{\scriptstyle\sqrt{2}} ~\mathrm{X}\2\,\ep\,\phi'_s\,, 	\label{s-1/2,s on} 
	\end{align}
	if one uses the equations of motion, i.e. $\mathrm{Y}\2\,\psi'_{s-1}=(-\,\ws{\scriptstyle/}{\scriptstyle\sqrt{\smash[b]{2(N+1)}}})\,\mathcal{F}\,\psi'_{s-1}=0$, and $h'_s=0$.

	\section{Conclusions and outlook}\label{conclu}

	In this work, we extended on-shell description \cite{Najafizadeh:2019mun} of the supersymmetric continuous spin gauge theory to an off-shell description. For this purpose, we first considered unconstrained formulation of the CSP theory \`a la Segal, given by the bosonic \cite{Schuster:2014hca} and the fermionic \cite{Najafizadeh:2015uxa} CSP actions. By introducing an auxiliary CSP field, we found an appropriate auxiliary action to construct the off-shell SUSY CSP action \eqref{off susy u action}. We observed that such auxiliary (non-dynamical) field should be a complex scalar continuous spin gauge field which accompany with the dynamical fields (the complex scalar and the Dirac CSP fields) constitute the $\mathcal{N}=1$ continuous spin supermultiplet. We then provided supersymmetry transformations \eqref{susy tr} leaving the off-shell SUSY CSP action invariant. It was shown that the algebra of supersymmetry transformations can be closed without using the equations of motion, up to some gauge transformations. 
	
	\vspace{.5cm}
	Afterwards, we took into account constrained formulation of the continuous spin gauge theory \`a la Fronsdal, described by the bosonic \cite{Metsaev:2016lhs} and the fermionic \cite{Metsaev:2017ytk} CSP actions. Introducing an appropriate auxiliary action, we could construct the off-shell SUSY CSP action \eqref{susy off const}. We then provided SUSY transformations \eqref{susy tr cons} demonstrating that the supersymmetry algebra can be closed off-shell, up to gauge transformation. 
	
	\vspace{.5cm}
	As one expects, by applying equations of motion, an off-shell description of the supersymmetry theory should recover on-shell results. Therefore, using the equations of motion, we showed off-shell SUSY transformations, both the unconstrained \eqref{susy tr} and the constrained \eqref{susy tr cons} ones, will reproduce the on-shell results presented in \cite{Najafizadeh:2019mun}.
	
	\vspace{.5cm}
	As one could see, we established both formulations (unconstrained and constrained) in terms of complex fields (instead of real fields) under which the SUSY actions and transformations found a compact form. In other words, according to complex fields, one finds that fewer calculations are needed to illustrate action invariance and check the closure of the SUSY algebra, while this would be a tedious task when dealing with real fields. Nevertheless, in order to reproduce higher spin results one needs to rephrase these formulations based on real fields (see Fig.\,\ref{fig1}). Thus, we reformulated constrained formulation of the SUSY CSP theory in terms of real fields. In this sense, the form of the off-shell SUSY CSP action and its SUSY transformations were respectively presented in \eqref{real csp action},\eqref{real susy tr}.
	\begin{figure}[t]
		\centering
		\includegraphics[scale=.20]{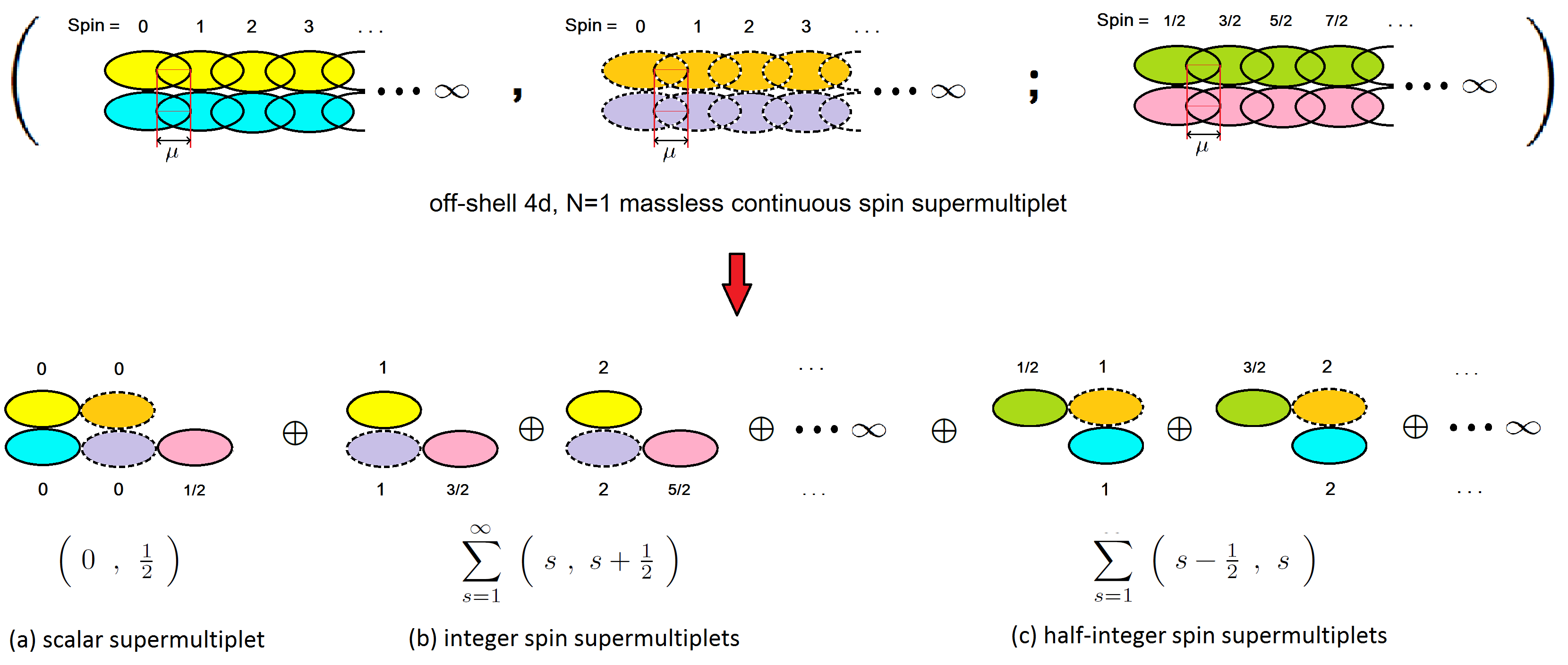}
		\caption{This figure illustrates, when $\m$ vanishes, the continuous spin supermultiplet, comprising of the complex scalar, auxiliary complex scalar, and Dirac continuous spin gauge fields, decomposes into a direct sum of all known off-shell helicity supermultiplets. Colored dashed ellipses represent auxiliary fields in each supermultiplet.}
		\label{fig3}
	\end{figure} 
	  	
	\vspace{.5cm}
	
	By taking the helicity limit ($\m\rightarrow0$), we derived the so-called ``off-shell supersymmetric higher spin action'' \`a la Fronsdal \eqref{real hs action}, in which the spin-$0$ field was included in the spectrum. The supersymmetry transformations were presented in \eqref{real susy tr hs}. Such supersymmetric higher spin theory was obtained from a top-down method, i.e. from the SUSY continuous spin theory in the limit $\m\rightarrow0$, in which the spin-$0$ field already existed in the spectrum (see e.g. \eqref{Phi, xi} or Fig.\,\ref{fig1}). This is why, along with all known integer and half-integer higher spin supermultiplets, we expected to reproduce the Wess-Zumino supermultiplet in particular.

	\vspace{.5cm}
	
	Therefore, for the above purpose, we decomposed the off-shell SUSY higher spin action \eqref{real hs action} into three parts; each one corresponds to the Wess-Zumino supermultiplet $(\,{\scriptstyle 0}\,,\, {\scriptstyle 1/2}\,)$, integer spin supermultiplets $(\,s\,,\, s\, {\scriptstyle +\, 1/2}\,),\, s\geqslant1$; and half-integer spin supermultiplets $(\, s\, {\scriptstyle -\, 1/2}\,,\, s \,\,),\, s\geqslant1$. For each part, we could then read corresponding SUSY transformations from \eqref{real susy tr hs}. In particular, for $s=2$, we reproduced off-shell 4d $\mathcal{N}=1$ linearized supergravity multiplet in flat spacetime. Our results can be visualized on Fig.\,\ref{fig3} demonstrating how the off-shell continuous spin supermultiplet decomposes into the all known off-shell 4d $\mathcal{N}=1$ supermultiplets, when $\m$ vanishes.

	\vspace{.5cm}

	Finally, let us conclude by briefly addressing some remarks and possible further studies. There is a formulation in which infinite sets of higher spin fields appear (see \cite{Sorokin:2017irs} for a review). Supersymmetric higher spin models constructed in hyperspace \cite{Bandos:1999qf,Bandos:2004nn,Florakis:2014kfa,Florakis:2014aaa} describe infinite-dimensional higher spin supermultiplets which are different from the conventional higher spin supermultiplets acquired in the helicity limit of the SUSY CSP. It is interesting to study on possible interacting SUSY CSPs either in on-shell or off-shell description. To this end, one may need to develop cubic interaction vertices of the 4d $\mathcal{N}=1$ arbitrary spin massless supermultiplets, see for example \cite{Metsaev:2019dqt, Metsaev:2019aig, Khabarov:2020bgr, Khabarov:2020deh, Gates:2019cnl} and references therein\,\footnote{In 3d flat spacetime, see for example \cite{Metsaev:2020gmb, Metsaev:2021bjh, Zinoviev:2021cmi}.}. In this work, we applied component approach to supersymmetrize the CSP theory off-shell. Therefore, it would be nice to establish a superspace formulation for the continuous spin gauge theory in which the helicity limit ($\m\rightarrow0$) should reproduce off-shell higher spin results \cite{Kuzenko:1993jp, Kuzenko:1993jq}. Finally, we note that recently an off-shell description for the 4d $\mathcal{N}=1$ massive supermultiplets was found for half-integer supermultiplets \cite{Koutrolikos:2020tel}\,\footnote{We note that on-shell description of integer and half-integer massive supermultiplets was first found in \cite{Zinoviev:2007js}.}. By constructing off-shell massive integer supermultiplets (which is still an open problem), one can obtain a direct sum of all 4d $\mathcal{N}=1$ massive supermultiplets. Then, if one takes the continuous spin limit (i.e. $m\rightarrow 0$\,, $s \rightarrow \infty$\, while $ms=\m=\mbox{constant}$), it is expected to be related to our results in this work.

	\section*{Acknowledgments}
	
	We are grateful to Hamid Afshar, Bernard de Wit, Abu Mohammad Khan, Konstantinos Koutrolikos, Sergei Kuzenko, Shahin Sheikh-Jabbari, and Dmitri Sorokin for useful discussions and comments. 
	
	
	\appendix
	
	\section
	{Conventions}\label{conven}
	
	We work in the 4-dimensional Minkowski spacetime and use the {\bf mostly plus} signature for the metric $\e_{\a\b}$. We denote coordinates with $x^\m$, auxiliary coordinates in unconstrained formulation with $\e^\m$, and auxiliary coordinates in constrained formulation with $\w^\m$. As a shorthand, we define conventions
	\begin{align}
		\p_\n:=\frac{\p}{\p x^\n}\,,   \qquad \eb_\n:=\frac{\p}{\p {\e^{\n}}}\,, \qquad \wb_\n:=\frac{\p}{\p {\w^{\n}}} \,,
		\qquad \Box:=\p^\n\p_\n\,, \qquad N:=\e^\n\,\eb_\n\,, \qquad N:=\w^\n\,\wb_\n\,,\label{def1}
	\end{align}
	and\,\footnote{Note that, in unconstrained formulation, an operator say $N:=\e\c\eb$ does not appear in flat space, it will appear in (A)dS space \cite{Najafizadeh:2018cpu}.} thus the following commutation relations
	\be
	\big[\,\eb^{\,\a} \,,\, \e^{\,\b}\,\big] =  {\eta}^{\,\a\b}\,,
	\quad \quad 
	\big[\, \eb^{\,2} \,,\, \e^{\,2}\,\big] =  4\, (N + 2) \,, 
	\quad \quad 
	\big[\,\wb^{\,\a} \,,\, \omega^{\,\b}\,\big] =  \eta^{\,\a\b}\,,  
	\quad \quad 
	\big[\,{\dww} \,,\, \omega^{\,2}\,\big] =  4\, (N + 2) \,, 
	\ee
	are used (for a full list of commutation relations see appendix D in \cite{Najafizadeh:2017acd}). The Hermitian conjugation rules in $\e$-space (unconstrained formulation \ref{unconstrained formu}) define as
	\be 
	(\p^{\,\n})^\dag:=\,-\,\p^{\,\n} \,,\qquad (\eb^{\,\n})^\dag := -\,\eb^{\,\n}\,,\qquad (\e^{\,\n})^\dag := \e^{\,\n}\,,\label{Hermitian con 1}
	\ee 
	and in $\w$-space (constrained formulation \ref{constrained formu}) introduce as 
	\be
	(\p^{\,\n})^\dag:=\,-\,\p^{\,\n} \,,\qquad (\dw^{\,\n})^\dag := \w^{\,\n} \,,\qquad (\w^{\,\n})^\dag:=\dw^{\,\n} \,.
	\label{hermitian conjugates 2}
	\ee
	%
	For the 4-dimensional Dirac gamma-matrices we use
	the conventions
	\be
	\big\{\,\g^\m \,,\,\g^\n \,\big\}=2\,\e^{\,\m\n}
	\,,
	\quad\quad\quad\quad
	(\,\g^\m\,)^{\,\dag} = \g^0\,\g^\m\,\g^0 \,,
	\quad\quad\quad\quad
	(\,\g^0\,)^{\,\dag} =-\, \g^0\,,
	\quad\quad\quad\quad
	(\,\g^0\,)^{\,2} =-\, 1\,,
	\label{anti comm gamma}
	\ee
	\be
	\big\{\,\gamma^\m \,,\, \gamma^5 \,\big\}=0 \,,~~\quad\quad\quad \quad\quad
	\g^5:=i\,\g^0\,\g^1\,\g^2\,\g^3\,,\quad\quad\quad \quad \quad
	(\,\g^5\,)^{\,\dag} = \g^5\,, \quad\quad\quad\quad\quad
	(\,\g^5\,)^{\,2}=1\,.
	\ee
	The Dirac adjoint defines as\,\footnote{In the mostly plus signature for the metric, the Dirac adjoint usually defines with an $i$, while in the mostly minus signature it defines as $\overline\Psi:=\Psi^\dagger\g^0$.}
	\be 
	\overline\Psi:=\Psi^\dagger\,i\,\g^0\,,\label{def2}
	\ee  
	and the Dirac slash notations as
	\be 
	\ds:=\g^\n\,\p_\n\,,\quad\quad\quad \es:=\g^\n\,\e_\n\,,  \quad\quad\quad \ebs:=\g^\n\,\eb_\n \,, \quad\quad\quad\ws:=\g^\n\,\w_\n\,, \quad\quad\quad \wbs:=\g^\n\,{\dw}_\n\,, \label{def3}
	\ee 
	thus, the anti-commutation relations become
	\be 
	\big\{\,\ebs \,,\, \es \,\big\} =2\,(N + {2}) \,, \qquad\qquad\qquad \big\{\,\wbs \,,\, \ws \,\big\} =2\,(N + {2})\,.
	\ee

	\section
	{SUSY action invariance}\label{invariance}
	This appendix is devoted to explicitly illustrate invariance of the supersymmetric action \eqref{off susy u action} under supersymmetry transformations \eqref{susy tr}. For this purpose, using Hermitian conjugation rules \eqref{Hermitian con 1}, one may calculate adjoint of supersymmetry transformations \eqref{susy tr} and conveniently arrive at
	\begin{subequations}
		\label{susy tr adj}
		\begin{align}
			&\delta\,\Phi^\dagger\, =\,-\,{\scriptstyle\sqrt{2}}\,\,\overbar{\Psi} \,\big(\es-1\,\big)\,R~\ep\,, \label{susy tr11}\\[8pt]
			&\delta\,\mathrm{H}^\dagger\, = \,+\,{\scriptstyle\sqrt{2}}\,\,\overbar{\Psi}\,\big(\g^0\,\Bbb{F}^\dagger\,\g^0\big) \,\big(\es+1\,\big)\,L~\ep \,,  \label{susy tr22}\\[8pt]
			&\,\delta\,\overbar{\Psi}\,\, =\,-\,{\scriptstyle\sqrt{2}}\,\,\Phi^\dagger \,\bar{\ep}\,L\,\big(\g^0\,\Bbb{X}^\dagger\,\g^0\big)\,-\, 
			{\scriptstyle\sqrt{2}}\,\,\mathrm{H}^\dagger\,\bar{\ep}\,R  \,, \label{susy tr33}
		\end{align}
	\end{subequations}
	in which $\Bbb{F}, \Bbb{X}, R, L$ were introduced in \eqref{operator F},\eqref{bbb X},\eqref{projections} respectively. Then, by varying the supersymmetric action \eqref{off susy u action} with respect to the fields, one gets
	\begin{align}
		\d\,S&=\int d^4x\, d^4\e\,\,\Big[\,\d\,\Phi^\dagger~\delta'(\e^2-1)\,\Bbb{B}~\Phi~+~\d\,\mathrm{H}^\dagger~\delta'(\e^2-1)\,\mathrm{H}~+~\d\,\overline\Psi\,\delta'(\e^2-1)\,(\es-1)\,\Bbb{F}~\Psi \nonumber\\[4pt]
		& \qquad \qquad\qquad~~~~ \Phi^\dagger~\delta'(\e^2-1)\,\Bbb{B}~\d\,\Phi~+~\mathrm{H}^\dagger~\delta'(\e^2-1)\,\d\,\mathrm{H}~+~\overline\Psi\,\delta'(\e^2-1)\,(\es-1)\,\Bbb{F}~\d\,\Psi  \,\Big] \,.\label{vary}
	\end{align} 
	Now, by plugging supersymmetry transformations \eqref{susy tr} and their adjoints \eqref{susy tr adj} into \eqref{vary}, one will arrive at
	\be 
	\d\,S=0\,.
	\ee 
	To obtain the latter, the following useful relations, which can be conveniently proved, may be used
	\begin{align} 
		&\Bbb{B}=\Bbb{F}\,\Bbb{X}\,,\\[4pt] 
		&\big(\g^0\,\Bbb{X}^\dagger\,\g^0\big)\,\delta'(\e^2-1)\,(\es-1\,)\,\Bbb{F}=\delta'(\e^2-1)\,\Bbb{B}\,(\es+1)\,,\\[4pt]
		&\big(\g^0\,\Bbb{F}^\dagger\,\g^0\big)\,\delta'(\e^2-1)\,(\es+1\,)=\delta'(\e^2-1)\,(\es-1\,)\,\Bbb{F}\,.
	\end{align}

	\section
	{Useful relations}\label{usef}
	
	The ``Majorana flip relations'' or the so-called ``Dirac flip relations'' are given by
	\begin{align}
		\bar\ep_2\,(\g^{\m_1}\,\g^{\m_2}\,\cdots\,\g^{\m_p})\,\ep_1&=(-1)^p~\bar\ep_1\,(\g^{\m_p}\,\cdots\,\g^{\m_2}\,\g^{\m_1})\,\ep_2 \,, \label{maj flip}
	\end{align}
	where $\ep_1$ and $\ep_2$ can be either the Majorana spinors or Dirac spinors depending on the problem we are dealing with (see \cite{Freedman:2012zz}, page 49, for more details).
	
	\vspace{.5cm}

	In order to illustrate that the SUSY CSP action is invariant under supersymmetry transformations, the following obtained relations are useful:
	\bea
	\hspace{-.75cm}\wdx\wbs(\ws\ds\wbs)&=&\wdx\ws\ds\dww-2\,\ws\wdx\dwdx\wbs+2\,\wdx \,h\, \ds\wbs\label{01}
	\\[5pt]
	\hspace{-.75cm}(\ws\ds\wbs)(\ws\ds\wbs)&=&-\,\w^2\,\Box\,\dww+2\,\ws\Box\, h\, \wbs+2\,\w^2\ds \dwdx\wbs+2\,\ws\wdx\ds\,\dww-4\,\ws\wdx\dwdx\wbs
	\\[5pt] 
	\hspace{-.75cm}\tfrac{1}{2}\,\ws\wdx\dww(\ws\ds\wbs)&=&-\,\ws\wdx\ds\dww+2\,\ws\wdx\dwdx\wbs+\tfrac{1}{2}\,\w^2\wdx\ds\wbs^3
	\\[5pt] 
	\hspace{-.75cm}\tfrac{1}{2}\,\w^2\dwdx\wbs(\ws\ds\wbs)&=&\tfrac{1}{2}\,\w^2\Box\dww-\w^2\ds\dwdx\wbs+\w^2\ds(h+1)\dwdx\wbs
	\\[5pt] 
	\hspace{-.75cm}-\,\tfrac{1}{4}\,\w^2\ds\dww(\ws\ds\wbs)&=&\tfrac{1}{2}\,\w^2\Box\dww-\w^2\ds\dwdx\wbs-\tfrac{1}{2}\,\w^2\wdx\ds\wbs^3
	\\[5pt]
	\hspace{-.75cm}(\ws\ds\wbs)(\ws\wbs)&=&\w^2\ds\dww-2\,\ws\wdx\dww+2\,\ws\ds h\wbs
	\\[5pt]
	\hspace{-.75cm}(\ws\ds\wbs)(\w^2\dww)&=&-2\,\w^2\ds\dww+4\,\ws\wdx\dww
	\\[5pt]
	\hspace{-.75cm}\tfrac{1}{2}\,\w^2\dwdx\wbs(\ws\wbs)&=&-\,\tfrac{1}{2}\,\w^2\ds\dww+\w^2(h+1)\dwdx\wbs
	\\[5pt]
	\hspace{-.75cm}\tfrac{1}{2}\,\w^2\dwdx\wbs(\w^2\dww)&=&\w^2\ds\dww+\w^2 \wdx\wbs^3
	\\[5pt] 
	\hspace{-.75cm}-\,\tfrac{1}{4}\,\w^2\ds\dww (\ws\wbs)&=&-\,\tfrac{1}{2}\,\w^2\wdx\wbs^3-\,\tfrac{1}{2}\,\w^2\ds\dww \label{10}
	\eea
	where $h:=\w\c\dw+\frac{d}{2}$\,, such that $d$ is spacetime dimension. We obtained these relations using (anti-)commutation relations presented in appendix of D in \cite{Najafizadeh:2017acd} and have dropped terms of order $\mathcal{O}(\w^3)$ and $\mathcal{O}(\dw^4)$ which vanish by constraints $\overline{\psi}(x,\dw)\,\ws^3=0$ and $(\dww)^2\,\phi(x,\w)=0$ in the SUSY action.

	\vspace{.5cm}
	
	In order to close the SUSY algebra, one can use the identity
	\be 
	\g^\m\g^\n\g^\r=\e^{\m\n}\g^\r+\g^\m \e^{\n\r}- \e^{\m\r}\g^\n - i\,\ep^{\a\m\n\r}\,\g_\a\,\g^5\,,
	\ee 
	leading to
	\be 
	\ws \ds\, \wbs = (\w\c\p)\,\wbs+\ws(\dw\c\p) - N\,\ds -i\,\ep^{\a\m\n\r}\,\w_\m\, \p_\n\, {\dw}_\r (\g_\a\,\g^5)\,,\label{123}
	\ee 
	where $N:=\w\c\dw$. Moreover, defining chiral projectors
	\be 
	L:=\tfrac{1}{2}\,(1+\g^5) \,, \qquad R:=\tfrac{1}{2}\,(1-\g^5)\,, 
	\ee  
	one can show
	\be 
	R\,(\ep\2\,\bar\ep\1-\ep\1\,\bar\ep\2)\,L=\tfrac{1}{2}\,\bar\ep\2\,\g_\n\,\ep\1\,\g^\n L\,,\qquad \qquad
	L\,(\ep\2\,\bar\ep\1-\ep\1\,\bar\ep\2)\,R=\tfrac{1}{2}\,\bar\ep\2\,\g_\n\,\ep\1\,\g^\n R\,,
	\ee
	where the following identity has been applied
	\be 
	\ep_2\,\bar\ep_1-\ep_1\,\bar\ep_2-\g^5\,(\ep_2\,\bar\ep_1-\ep_1\,\bar\ep_2)\,\g^5=-\,\bar\ep_1\,\g_\m\,\ep_2\,\g^\m\,.
	\ee

	\setstretch{0.95}

\end{document}